\documentclass[journal]{IEEEtran}
\usepackage{stfloats}
\usepackage{cite}
\usepackage{float}
\usepackage{graphicx}
\usepackage{subfigure}
\usepackage{color}
\usepackage{amssymb}
\usepackage{amsmath}
\usepackage{algorithm}
\usepackage{algorithmic}
\usepackage{bm}
\usepackage{hyperref}

\begin{document}

\title{Average Secrecy Capacity Maximization of Rotatable Antenna-Assisted Secure Communications}

\author{Pengchuan Jiang, Quanzhong Li, Lifeng Mai, and Qi Zhang, \emph{Member}, \emph{IEEE}

\thanks{This work has been submitted to the IEEE for possible publication. Copyright may be transferred without notice, after which this version may no longer be accessible.

\emph{(Corresponding author: Qi Zhang.)}

Pengchuan Jiang and Qi Zhang are with the School of Electronics and Information Technology, Sun Yat-sen University, Guangzhou 510006, China (e-mail: jiangpch5@mail2.sysu.edu.cn; zhqi26@mail.sysu.edu.cn). Quanzhong Li is with the School of Computer Science and Engineering, Sun Yat-sen University, Guangzhou 510006, China (e-mail: liquanzh@mail.sysu.edu.cn). Lifeng Mai is with the Electric Power Research Institute, China Southern Power Grid, Guangzhou 510663, China, and also with Guangdong Provincial Key Laboratory of Power System Network Security, Guangzhou 510663, China (e-mail:
mailf@csg.cn).}

}

\maketitle

\begin{abstract}
A rotatable antenna, which is able to dynamically adjust its deflection angle, is promising to achieve better physical layer security performance for wireless communications. In this paper, considering practical scenarios with non-real-time rotatable antenna adjustment, we investigate the average secrecy rate maximization problem of a rotatable antenna-assisted secure communication system. We theoretically prove that the objective function of the average secrecy rate maximization problem is quasi-concave with respect to an adjustment factor of the rotatable antenna. Under this condition, the optimal solution can be found by the bisection search. Furthermore, we derive the closed-form optimal deflection angle for the secrecy capacity maximization problem, considering the existence of only line-of-sight components of wireless channels. This solution serves as a near optimal solution to the average secrecy rate maximization problem. Based on the closed-form near optimal solution, we obtain the system secrecy outage probability at high signal-to-noise ratio (SNR). It is shown through simulation results that the near optimal solution achieves almost the same average secrecy capacity as the optimal solution. It is also found that at high SNR, the theoretical secrecy outage probabilities match the simulation ones.
\end{abstract}


\section{Introduction}

Physical layer security is a promising technique to ensure security and privacy in future wireless communications. To enhance physical layer security performance, advanced fluid antenna system (FAS) \cite{KKWong21,TWu24,TWu25,JZheng24} and movable antenna system (MAS) \cite{LZhu24,XShao25,XShao252} were exploited to optimize channel conditions. It is demonstrated in \cite{Ghadi24,Sanchez24} that FAS achieves higher secrecy rate and secrecy outage performance than the conventional multi-antenna systems by using spatial degrees of freedom. Furthermore, by jointly optimizing beamforming and antenna positions, FAS significantly enhances secrecy rate while satisfying covertness constraints \cite{JYao25}.  In \cite{GHu24,JTang25,JDing25}, MAS based physical layer security schemes were proposed for point-to-point, multi-user, and full-duplex multiple-input multiple-output systems. 

To further exploit spatial degrees of freedom, the concept of a six-dimensional movable antenna (6DMA) system was proposed. The system enables independent adjustment of both the three-dimensional (3D) positions and 3D rotation angles of antenna surfaces \cite{XShao25,XShao252}, thereby opening a new paradigm for achieving better physical layer security performance. In \cite{YQian}, the potential of 6DMA for enhancing physical layer security performance was investigated.

Although 6DMA provides a general model for position and rotation adjustable antennas, the properties of its rotation adjustable antennas were not fully investigated. To address this issue, Zheng \emph{et al.} focus on rotatable antennas and propose a rotatable antenna enabled wireless communication system \cite{BZheng25,BZheng252,QWu25}. In the system, the transmitter is equipped with a flexible antenna that dynamically adjusts its deflection angle through mechanical or electronic means to reconstruct the beam direction in 3D space. The rotatable antenna technique has shown its potential in various communication scenarios. In \cite{RZhao25}, Zhao \emph{et al.} proposed a movable and rotatable antenna that adjusts both its position and deflection angle. In \cite{XXiong25}, the rotatable antenna was exploited to sense the environment from different directions, thereby improving channel state information accuracy. In \cite{XPeng25}, the application of rotatable antenna on spectrum-sharing was studied. In \cite{XZhang25}, it was found that deploying rotatable antenna arrays on unmanned aerial vehicle (UAV) base station (BS) enables the channel vectors of different users to become asymptotically orthogonal, thereby reducing inter-user interference and enhancing transmission efficiency. In \cite{CZhou25}, it was shown that rotatable antennas for integrated sensing and communications (ISAC) enhance communication rates while simultaneously reducing the Cram\'er-Rao bound of sensing performance.

Regarding rotatable antenna-assisted secure communications, Dai \emph{et al.} proposed to maximize the secrecy rate by jointly optimizing transmit beamforming and the deflection angles of multiple rotatable antennas \cite{LDai25}. In \cite{LDai25}, the global channel state information (CSI) is assumed to be known. Furthermore, it is assumed that the
duration of channel estimation and rotatable antenna adjustment followed by secure message transmission is shorter than the coherent time of wireless channel. The assumption may be stringent for practical rotatable antenna design. 

In this paper, we investigate the average secrecy rate maximization problem of a rotatable antenna-assisted secure communication system. In the system, the instantaneous CSI is not required to be known. Furthermore, the adjustment time of rotatable antenna is not a stringent constraint. Our main contributions are summarized as follows:

\begin{itemize}
\item We theoretically prove that to maximize the average secrecy rate, the optimal boresight vector for the rotatable antenna should intersect the extension line from the eavesdropper's position to the legitimate user's position. Based on this finding, we theoretically prove that the objective function of the average secrecy rate maximization problem is quasi-concave with respect to an adjustment factor of the rotatable antenna. Under this condition, the optimal solution can be found by the bisection search.  
    
\item The optimal solution obtained by the bisection search is a numerical solution. We propose to solve the secrecy capacity maximization problem considering the existence of only line-of-sight (LoS) components of wireless channels, which serves as a near optimal solution to the average secrecy rate maximization problem. Considering the existence of only LoS components, we theoretically derive the closed-form optimal deflection angle. 
    
\item At high signal-to-noise ratio (SNR), based on the closed-form near optimal solution, we theoretically derive the system secrecy outage probability. The theoretically derived system secrecy outage probability, which is not related to the deflection angle of rotatable antenna, provides us a secrecy outage probability upper bound.

\end{itemize}

The remainder of this paper is organized as follows. Section II presents the model of a rotatable antenna-assisted secure communication system. Section III and Section IV derive the optimal and closed-form near optimal solutions, respectively, for average secrecy capacity maximization. Section V obtains the closed-form secrecy outage probability at high SNR. Section VI presents the simulation results for system performance evaluation. Finally, Section VII concludes this paper.

\emph{Notations}: Boldface lowercase and uppercase letters denote vectors and
matrices, respectively. The $\mathbf{A}^T$ denotes the transpose of the matrix $\mathbf{A}^T$. The vector from point $A$ to point $B$ is denoted as $\overline{AB}$.

\section{System Model}

Consider a rotatable antenna-assisted secure communication system, where a BS, equipped with a single rotatable antenna, transmits confidential information to a legitimate user, while an eavesdropper attempts to intercept. Both the legitimate user and the eavesdropper are equipped with a single isotropic fixed antenna. To clearly describe the rotatable antenna, a 3D Cartesian coordinate system is established, where the rotatable antenna is positioned at the origin. Relative to the rotatable antenna, the position vectors of legitimate user and eavesdropper are denoted as
\begin{align}
\mathbf{q}_b=&[x_b, y_b, z_b]^T,\\
\mathbf{q}_e=&[x_e, y_e, z_e]^T,
\end{align}
respectively. Let
\begin{equation}
\mathbf{q}_a(\bm{\theta})=\|\mathbf{q}_a(\bm{\theta})\|\left[\sin\theta_z\cos\theta_a,\sin\theta_z\sin\theta_a,\cos\theta_z\right]^T
\end{equation}
be the 3D boresight vector of the rotatable antenna, where
\begin{equation}\label{q4}
\bm{\theta}\triangleq[\theta_z,\theta_a]^T
\end{equation}
denotes its deflection angle vector. In \eqref{q4}, the zenith angle $\theta_z$ represents the angle between the boresight direction and z-axis, and the azimuth angle $\theta_a$ denotes the angle between the projection of the boresight direction onto the x-y plane and x-axis \cite{BZheng25}.

According to \cite{LDai25}, the channels from BS to the legitimate user and eavesdropper, denoted as $h_b(\bm{\theta})$ and $h_e(\bm{\theta})$, respectively,
are modeled as quasi-static flat-fading channels
\begin{align}\label{q5}
h_i(\bm{\theta})=\sqrt{L_iG_{i}(\bm{\theta})}u_{i}
\end{align}
for $i\in\{b,e\}$, where $L_i$ denotes the large-scale fading component, $G_{i}(\bm{\theta})$ denotes the effective antenna gain for rotatable antenna, and
$u_{i}$ represents the small-scale fading component. In \eqref{q5}, the large-scale fading component $L_i$ is modeled as $L_i=\zeta_0\|\mathbf{q}_i\|^{-\beta_i}$, where $\zeta_0$ denotes the channel power gain at the reference distance of one meter and $\beta_i$ denotes the path-loss exponent. From \cite{BZheng25,LDai25}, the effective antenna gain for
rotatable antenna is expressed as
\begin{equation}\label{q6}
G_i(\bm{\theta})=\left\{\begin{array}{lc}
G_0\cos\epsilon_i; & \epsilon_i\in[0,\pi/2] \\
0; & \text{otherwise} 
\end{array}
\right.
\end{equation}
where $G_{0}=4$ denotes the maximum gain in the boresight direction, and $\epsilon_i$ denotes the angle between $\mathbf{q}_a(\bm{\theta})$ and $\mathbf{q}_i$. Thus, we have
\begin{equation}\label{q7}
\cos\epsilon_i=\frac{\mathbf{q}^T_a(\bm{\theta})\mathbf{q}_i}{\|\mathbf{q}_a(\bm{\theta})\|\|\mathbf{q}_i\|}.
\end{equation}
The small-scale fading component $u_{i}$ is modeled as
\begin{align}\label{q8} u_{i}=\sqrt{\frac{K_i}{K_i+1}}u^{\text{LoS}}_i+\sqrt{\frac{1}{K_i+1}}u^{\text{NLoS}}_i
\end{align}
where $K_i$ denotes the Rician factor,
\begin{equation}
u^{\text{LoS}}_i=\exp(-j2\pi \|\mathbf{q}_i\|/\lambda)
\end{equation}
denotes the LoS component, and $u^{\text{NLoS}}_i\sim\mathcal{CN}(0,1)$ denotes the non-line-of-sight (NLoS) channel component.

The instantaneous SNRs at legitimate user and eavesdropper, denoted as $\gamma_b(\bm{\theta})$ and $\gamma_e(\bm{\theta})$, respectively, are given by
\begin{align}
\gamma_i(\bm{\theta})=\frac{P|h_i(\bm{\theta})|^2}{\sigma^{2}}=\gamma|h_i(\bm{\theta})|^2
\end{align}
where $P$ denotes the transmission power of the BS and $\gamma=P/\sigma^{2}$ denotes the SNR. The secrecy capacity is given by
\begin{equation}\label{q11}
C_s(\bm{\theta})=\left[C_b(\bm{\theta})-C_e(\bm{\theta})\right]^+
\end{equation}
where $[a]^+=\max\{a,0\}$ and
\begin{align}\label{q12}
C_i(\bm{\theta})=\log_2(1+\gamma_i(\bm{\theta}))
\end{align}
for $i\in\{b,e\}$.

In this paper, we consider that the 3D boresight vector of the rotatable antenna, $\mathbf{q}_a(\bm{\theta})$ cannot change as fast as the change of wireless channel. However, $\mathbf{q}_a(\bm{\theta})$ may change as fast as the position variations of legitimate user and eavesdropper. Therefore, provided with the positions of legitimate user and eavesdropper, our goal is to optimize $\mathbf{q}_a(\bm{\theta})$ which maximizes the average secrecy capacity over a relatively long period and analyze its secrecy outage probability.

\section{Optimal Solution for Average Secrecy Capacity Maximization}

Given the positions of legitimate user and eavesdropper, $\mathbf{q}_b$ and $\mathbf{q}_e$, the average secrecy capacity maximization problem is formulated as follows
\begin{align}\label{q13}
\max_{\bm{\theta}}\quad &\mathbb{E}[C_s(\bm{\theta})].
\end{align}
To continue, we have the following proposition.

\emph{Proposition 1}: The optimal boresight vector for the rotatable antenna, denoted as $\mathbf{q}_a^o(\bm{\theta})$, should intersect the extension line from the eavesdropper's position to the legitimate user's position.

\emph{Proof}: From \eqref{q5}, we know that the effect of $\bm{\theta}$ on the objective function of problem \eqref{q13}, i.e., $\mathbb{E}[C_s(\bm{\theta})]$, is due to $G_{b}(\bm{\theta})$ and $G_{e}(\bm{\theta})$. Maximizing $\mathbb{E}[C_s(\bm{\theta})]$ is equivalent to simultaneously maximizing  $G_{b}(\bm{\theta})$ and minimizing $G_{e}(\bm{\theta})$.

As shown in Fig. 1, we denote the positions of the BS, legitimate user, and eavesdropper as $A$, $B$, and $E$. The optimal boresight vector for the rotatable antenna $\mathbf{q}_a(\bm{\theta})$ lies on a cone with vertex $A$, axis $\overline{AB}$, and radius $\|\mathbf{q}_b\|\tan(\phi)$, where $\phi$ denotes the angle between $\mathbf{q}_a(\bm{\theta})$ and $\overline{AB}$. This is because the boresight vector on the cone guarantees the same $G_{b}(\bm{\theta})$. Maximizing $\mathbb{E}[C_s(\bm{\theta})]$ is equivalent to minimizing $G_{e}(\bm{\theta})$ under the constraint that the boresight vector is on the cone.

\begin{figure}
\centering
\includegraphics[width=2.5in]{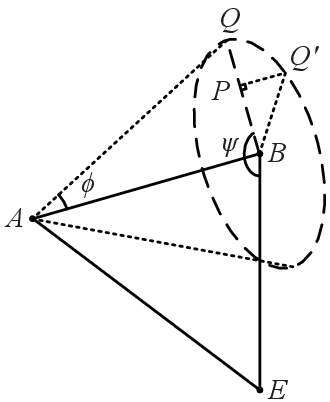}
\caption{Illustration of a rotatable antenna-assisted secure communication system.}
\end{figure}

From \eqref{q5} and \eqref{q7}, minimizing $G_{e}(\bm{\theta})$ is equivalent to finding the optimal boresight vector whose projection onto the vector $\mathbf{q}_e$ is minimized. Denote by $\Gamma$ the plane passing through points $A$, $B$, and $E$. The intersection of the cone and plane $\Gamma$ generates two boresight vectors, with the one farther from point $E$ denoted as $\overline{AQ}$.

It is noted that the vector $\overline{AQ}$ intersects the extension line from the eavesdropper's position to the legitimate user's position. We now prove that $\overline{AQ}$ is the optimal boresight vector for the rotatable antenna by contradiction. Suppose that the optimal boresight vector is $\overline{AQ'}$ which lies on the cone and is geometrically distinct from $\overline{AQ}$. Since the optimal boresight vector should satisfy that its projection onto the vector $\mathbf{q}_e$ is minimized, by contradiction, we will show that the projection of $\overline{AQ'}$ onto $\mathbf{q}_e$ is larger than that of $\overline{AQ}$.

From the law of cosines, we know
\begin{align}
\left\|\mathbf{q}_a(\bm{\theta})-\mathbf{q}_e\right\|^2=\left\|\mathbf{q}_a(\bm{\theta})\right\|^2+\left\|\mathbf{q}_e\right\|^2-2\mathbf{q}^T_a(\bm{\theta})\mathbf{q}_e.
\end{align}
To show that the projection of $\overline{AQ'}$ onto $\mathbf{q}_e$ is larger than that of $\overline{AQ}$ is equivalent to proving
\begin{align}\label{q15}
\left\|\overline{EQ'}\right\|<\left\|\overline{EQ}\right\|.
\end{align}

From Fig. 1, we have
\begin{align}
\left\|\overline{EQ'}\right\|^2=\left\|\overline{EP}\right\|^2+\left\|\overline{PQ'}\right\|^2.
\end{align}
Using the law of cosines, we know
\begin{align}\label{q17}
\left\|\overline{EP}\right\|^2&=\left\|\overline{EB}\right\|^2+\left\|\overline{BP}\right\|^2-2\left\|\overline{EB}\right\|\left\|\overline{BP}\right\|\cos\psi
\end{align}
where $\psi$ denotes the angle between $\overline{BE}$ and $\overline{BP}$. From \eqref{q17}, $\|\overline{EP}\|$ is maximized and thus $\left\|\overline{EQ'}\right\|$ is maximized when $\psi\in[\pi/2,3\pi/2]$, i.e., $\cos\psi<0$. Under this condition, $\psi$ is also the angle between $\overline{BE}$ and $\overline{BQ}$. Using \eqref{q17}, we obtain
\begin{align}\label{q18}
\left\|\overline{EQ'}\right\|^2=\left\|\overline{EB}\right\|^2+\left\|\overline{BQ'}\right\|^2-2\left\|\overline{EB}\right\|\left\|\overline{BP}\right\|\cos\psi.
\end{align}
Similarly, we obtain
\begin{align}\label{q19}
\left\|\overline{EQ}\right\|^2=\left\|\overline{EB}\right\|^2+\left\|\overline{BQ}\right\|^2-2\left\|\overline{EB}\right\|\left\|\overline{BQ}\right\|\cos\psi.
\end{align}
Comparing \eqref{q18} with \eqref{q19}, because $\|\overline{BQ'}\|=\|\overline{BQ}\|$, $\|\overline{BQ}\|>\|\overline{BP}\|$, and $\cos\psi<0$, we have \eqref{q15}. This contradicts that the optimal boresight vector is $\overline{AQ'}$. $\hfill\blacksquare$

Using Proposition 1, the optimal boresight vector can be expressed as
\begin{align}\label{q20}
\mathbf{q}_a(\bm{\theta})=\mathbf{q}_e+\alpha(\mathbf{q}_b-\mathbf{q}_e)
\end{align}
where $\alpha\geq1$ denotes an adjustment factor which establishes a bijective mapping with $\bm{\theta}$. When $\alpha=1$, $\mathbf{q}_a(\bm{\theta})=\mathbf{q}_b$ and when $\alpha \to \infty$, $\mathbf{q}_a(\bm{\theta})/\|\mathbf{q}_a(\bm{\theta})\|\rightarrow (\mathbf{q}_b-\mathbf{q}_e)/\|\mathbf{q}_b-\mathbf{q}_e\|$.

Given the adjustment factor $\alpha$, the optimal deflection angle vector of the rotatable antenna is computed as follows
\begin{align}
\theta_{z}&=\arccos\frac{z_e+\alpha(z_b-z_e)}{\|\mathbf{q}_e+\alpha(\mathbf{q}_b-\mathbf{q}_e)\|},\\ 	\theta_{a}&=\arctan\frac{y_e+\alpha(y_b-y_e)}{x_e+\alpha(x_b-x_e)}.
\end{align}

Therefore, the optimization problem \eqref{q13} is reduced to
\begin{align}\label{q23}
\max_{\alpha\geq1}\ \mathbb{E}[C_s(\alpha)].
\end{align}
It is noted that when
$\mathbf{q}^T_a(\bm{\theta})\mathbf{q}_e=0$, from \eqref{q7}, we have $G_e(\bm{\theta})=0$ and thus $\mathbb{E}[C_e(\bm{\theta})]$ is minimized, where the corresponding $\alpha$ is denoted as $\alpha_{\max}$. If $\alpha>\alpha_{\max}$, with the increase of $\alpha$, $\mathbb{E}[C_e(\bm{\theta})]$ remains 0 and $\mathbb{E}[C_b(\bm{\theta})]$ decreases. Therefore, the optimal $\alpha$ which maximizes $\mathbb{E}[C_s(\bm{\theta})]$ is over $[1,\alpha_{\max}]$.

Using \eqref{q20}, we have
\begin{equation}\label{q24}
\left(\mathbf{q}_e+\alpha_{\max}(\mathbf{q}_b-\mathbf{q}_e)\right)^T\mathbf{q}_e=0.
\end{equation}
Solving the equation \eqref{q24}, we obtain
\begin{equation}\label{q25}
\alpha_{\max}=\frac{\|\mathbf{q}_e\|^2}{\|\mathbf{q}_e\|^2-\mathbf{q}_b^T\mathbf{q}_e}.
\end{equation}
Problem \eqref{q23} is equivalent to
\begin{align}\label{q26}
\max_{1\leq\alpha\leq\alpha_{\max}}\ \mathbb{E}[C_s(\alpha)].
\end{align}
To proceed, we have the following proposition.

\emph{Proposition 2}: The objective function of problem \eqref{q23} $\mathbb{E}[C_s(\alpha)]$ is quasi-concave with respect to $\alpha$ over $1\leq\alpha\leq\alpha_{\max}$.

\emph{Proof}: From \eqref{q11} and \eqref{q12}, we know
\begin{align}\label{bq1}
\mathbb{E}[C_s(\alpha)]=\int_0^\infty\int_0^x\log_2\left(\frac{1+\gamma x}{1+\gamma y}\right)f_b(x;\alpha)f_e(y;\alpha)dydx
\end{align}
where $f_b(x;\alpha)$ and $f_e(y;\alpha)$ denote the probability density functions (PDFs) of the instantaneous channel power gains $|h_b(\alpha)|^2$ and $|h_e(\alpha)|^2$, respectively. From \eqref{q8}, $h_b(\alpha)$ and $h_e(\alpha)$ are Rician distributed random variables. Thus, $|h_b(\alpha)|^2$ and $|h_e(\alpha)|^2$ are noncentral chi-square distributed random variables with 2 degrees of freedom. Because the average channel power gain of $h_i(\alpha)$ is
\begin{equation}\label{bq2}
\mathbb{E}[|h_i(\alpha)|^2]=L_iG_0\Phi_i^{-1}(\alpha)
\end{equation}
for $i\in\{b,e\}$, where
\begin{align}\label{bq3}
\Phi_i^{-1}(\alpha)=\frac{(\mathbf{q}_e+\alpha(\mathbf{q}_b-\mathbf{q}_e))^T\mathbf{q}_i}{\|\mathbf{q}_e+\alpha(\mathbf{q}_b-\mathbf{q}_e)\| \|\mathbf{q}_i\|}.
\end{align}
The PDFs of $|h_b(\alpha)|^2$ and $|h_e(\alpha)|^2$ are \cite{Bhargav}
\begin{align}\label{bq4}
f_i(x;\alpha)=&\frac{\eta_i\Phi_i(\alpha)}{\exp(K_i)}\exp\left(-\eta_i\Phi_i(\alpha)x\right)\nonumber\\ &\cdot I_{0}\left(2\sqrt{K_i\eta_i\Phi_i(\alpha)x}\right)
\end{align}
for $i\in\{b,e\}$, where $I_{0}(x)=\frac{1}{\pi}\int_0^{\pi}\exp(x\cos t)dt$ denotes the zeroth-order modified Bessel function of the first kind and $\eta_i$ is defined as
\begin{align}\label{bq5}
\eta_i=&\frac{1+K_i}{L_iG_0}.
\end{align}
Substituting \eqref{bq4} and \eqref{bq5} into \eqref{bq1}, we have
\begin{align}\label{bq6}
\mathbb{E}&[C_s(\alpha)]=\frac{\exp(-K_b-K_e)}{\ln2}\int_0^\infty\int_0^x\Omega dydx
\end{align}
where
\begin{align}\label{bq7}
\Omega=&\eta_b\Phi_b(\alpha)\exp(-\eta_b\Phi_b(\alpha) x) I_{0}\left(2\sqrt{K_b\eta_b\Phi_b(\alpha) x}\right)\nonumber\\
\cdot&\eta_e\Phi_e(\alpha)\exp(-\eta_e\Phi_e(\alpha) y) I_{0}\left(2\sqrt{K_e\eta_e\Phi_e(\alpha) y}\right)\nonumber\\
\cdot&\ln\left(\frac{1+\gamma x}{1+\gamma y}\right).
\end{align}
In this paper, we prove Proposition 2 by showing that the integrand $\Omega$ is quasi-concave with respect to $\alpha$. From \eqref{bq7}, we know that $\alpha$ is involved in six terms, which complicates the proof of the quasi-concavity of $\Omega$.

Let
\begin{equation}
x'=\Phi_b(\alpha) x\text{ and }
y'=\Phi_b(\alpha) y.
\end{equation}
The expression of $\mathbb{E}[C_s(\alpha)]$ is modified as
\begin{align}\label{bq10}
\mathbb{E}&[C_s(\alpha)]=\frac{\exp(-K_b-K_e)}{\ln2}\int_0^\infty\int_0^{x'}\Omega' dy'dx'
\end{align}
where
\begin{align}
\Omega'=&\eta_b\exp(-\eta_bx') I_{0}\left(2\sqrt{K_b\eta_b x'}\right)\ln\left(\frac{\Phi_b(\alpha)+\gamma x'}{\Phi_b(\alpha)+\gamma y'}\right)\nonumber\\
\cdot&\frac{\eta_e\Phi_e(\alpha)}{\Phi_b(\alpha)}\exp\left(-\frac{\eta_e\Phi_e(\alpha)y'}{\Phi_b(\alpha)}\right) I_{0}\left(2\sqrt{\frac{K_e\eta_e\Phi_e(\alpha)y'}{\Phi_b(\alpha)}}\right).
\end{align}
The quasi-concavity of $\Omega'$ is equivalent to the quasi-concavity of $\ln\Omega'$ since the logarithm function is a monotonically increasing function. To show the quasi-concavity of $\ln\Omega'$, we compute the first-order partial derivative of $\ln\Omega'$ with respect to $\alpha$, which is expressed as follows
\begin{align}
\frac{\partial\ln \Omega'}{\partial \alpha}&=\frac{\partial \Lambda_1}{\partial \alpha}+\frac{\partial \Lambda_2}{\partial \alpha}
\end{align}
where
\begin{align}
\Lambda_1=&\ln\left(\ln\frac{\Phi_b(\alpha)+\gamma x'}{\Phi_b(\alpha)+\gamma y'}\right),\\
\Lambda_2=&\ln\frac{\eta_e\Phi_e(\alpha)}{\Phi_b(\alpha)}-\frac{\eta_e\Phi_e(\alpha)y'}{\Phi_b(\alpha)}\nonumber\\
&+\ln I_{0}\left(2\sqrt{\frac{K_e\eta_e\Phi_e(\alpha)y'}{\Phi_b(\alpha)}}\right).
\end{align}

It is noted that $\Lambda_1$ and $\Lambda_2$ are related to $\alpha$ via $\Phi_b(\alpha)$ and $\Phi_e(\alpha)$. From \eqref{bq3}, $\Phi_i(\alpha)$ is a monotonically increasing function with respect to $\alpha$, within the domain $\alpha\geq1$. To show the quasi-concavity of $\ln\Omega'$, the following proposition provides the relationship between $\Lambda_1$ and $\Phi_b(\alpha)$.

\emph{Proposition 3}: The function $\frac{\partial \Lambda_1}{\partial z}$ is a monotonically increasing function with respect to $z=\Phi_b(\alpha)$, whose lower and upper bounds are $-1$ and $0$, respectively. The lower and upper bounds are achieved when $z=1$ and $z\rightarrow\infty$, respectively.

\emph{Proof}: See Appendix A. $\hfill\blacksquare$

To analyze the relationship between $\Lambda_2$ and $\Phi_i(\alpha)$, we define
\begin{equation}\label{bq16}
\Psi=\frac{\Phi_e(\alpha)}{\Phi_b(\alpha)}.
\end{equation}
Substituting \eqref{bq3} into \eqref{bq16}, we have
\begin{equation}\label{bq17}
\Psi=\frac{\|\mathbf{q}_e\|}{\|\mathbf{q}_b\|}\cdot\frac{\mathbf{q}_e^T\mathbf{q}_b+\alpha\bar{\mathbf{q}}^T\mathbf{q}_b}{\|\mathbf{q}_e\|^2+\alpha\bar{\mathbf{q}}^T\mathbf{q}_e}
\end{equation}
where
\begin{equation}\label{bq18}
\bar{\mathbf{q}}=\mathbf{q}_b-\mathbf{q}_e.
\end{equation}
Taking the first-order partial derivative of $\Psi$ with respect to $\alpha$ yields
\begin{equation}\label{bq19}
\frac{\partial\Psi}{\partial \alpha}=\frac{\|\mathbf{q}_e\|}{\|\mathbf{q}_b\|}\cdot\frac{\bar{\mathbf{q}}^T\mathbf{q}_b\|\mathbf{q}_e\|^2-\bar{\mathbf{q}}^T\mathbf{q}_e\mathbf{q}_e^T\mathbf{q}_b}{(\|\mathbf{q}_e\|^2+\alpha\bar{\mathbf{q}}^T\mathbf{q}_e)^2}.
\end{equation}
To proceed, we need the following proposition.

\emph{Proposition 4}: The expression $\Psi$ in \eqref{bq16} is a monotonically increasing function with respect to $\alpha$.

\emph{Proof}: See Appendix B. $\hfill\blacksquare$

Using Proposition 4, instead of analyzing the relationship between $\Lambda_2$ and $\alpha$, we analyze the relationship between $\Lambda_2$ and $\Psi$. Taking the first-order partial derivative of $\Lambda_2$ with respect to $\Psi$ yields
\begin{align}\label{bq20}
\frac{\partial \Lambda_2}{\partial \Psi} &=\frac{1}{\Psi}\left(1+\sqrt{K_e\eta_e \Psi y'}\frac{I_1(2\sqrt{K_e\eta_e \Psi y'})}{I_0(2\sqrt{K_e\eta_e \Psi y'})}\right)-\eta_ey'
\end{align}
where $I_{1}(x)=\frac{1}{\pi}\int_0^{\pi}\exp(x\cos t)\cos t dt$ denotes the first-order modified Bessel function of the first kind. Because $\alpha\geq1$, from \eqref{bq17}, we have
\begin{equation}\label{bq21}
\Psi|_{\alpha=1}=\left(\frac{\mathbf{q}_e^T\mathbf{q}_b}{\|\mathbf{q}_e\|\|\mathbf{q}_b\|}\right)^{-1}=\cos^{-1}\angle BAE,
\end{equation}
whose lower and upper bounds are -1 and 1, respectively. Because $I_0(2\sqrt{K_e\eta_e \Psi y'})\geq0$ and $I_1(2\sqrt{K_e\eta_e \Psi y'})\geq0$, if $\Psi|_{\alpha=1}<0$, we have
\begin{equation}
\left.\frac{\partial \Lambda_2}{\partial \Psi}\right|_{\alpha=1}<0.
\end{equation}
If $\Psi|_{\alpha=1}>0$, the sign of $\frac{\partial \Lambda_2}{\partial \Psi}$ is not determined.

From \cite[Eqn. 9.7.1]{Abramowitz}, we know
\begin{align}
\lim_{x\rightarrow\infty}I_{\nu}(x)=\frac{\exp(x)}{\sqrt{2\pi x}}
\end{align}
for $\nu\in\{0,1\}$ and thus
\begin{align}\label{bq23}
\lim_{\Psi\rightarrow\infty}\frac{\partial \Lambda_2}{\partial \Psi}=-\eta_ey'\leq0.
\end{align}
To continue, we need the following propositions.

\emph{Proposition 5}: The function $\frac{\partial \Lambda_2}{\partial \Psi}$ is a monotonically decreasing function with respect to $\Psi$. If $\Psi|_{\alpha=1}\geq0$, $\frac{\partial \Lambda_2}{\partial \Psi}$ possesses a unique zero. Otherwise, it has no zero.

\emph{Proof}: See Appendix C. $\hfill\blacksquare$

\emph{Proposition 6}: Taking the first-order partial derivative of $z$ with respect to $\Psi$ yields
\begin{equation}\label{bq25}
\frac{\partial z}{\partial \Psi}=\frac{\|\mathbf{q}_e\|}{\|\mathbf{q}_a(\theta)\|}\cdot\frac{\cos^2\angle QAE}{\cos^2\angle QAB}(\alpha-1)\geq0.
\end{equation}

\emph{Proof}: See Appendix D. $\hfill\blacksquare$

Combining Proposition 3, Proposition 5, and Proposition 6, if $\frac{\partial \Lambda_2}{\partial \Psi}<0$ when $\alpha=1$, we know that
\begin{align}
\frac{\partial \ln\Omega'}{\partial \Psi}&=\frac{\partial \Lambda_1}{\partial z}\cdot\frac{\partial z}{\partial \Psi}+\frac{\partial \Lambda_2}{\partial \Psi}<0
\end{align}
has no zero. The optimal value of $\Psi$ which maximizes $\ln\Omega'$ is \eqref{bq21}. 
Therefore, $\Omega'$ and $\Omega$ are quasi-concave with respect to $\alpha$ over $1\leq\alpha\leq\alpha_{\max}$.

Otherwise, $\frac{\partial \Omega'}{\partial \Psi}$ possesses a unique zero. This can be verified as follows. From \eqref{bq25}, we have
\begin{align}
\left.\frac{\partial \Lambda_1}{\partial z}\cdot\frac{\partial z}{\partial \Psi}\right|_{\alpha=1}=0.
\end{align}
When $\alpha=\alpha_{\max}$, we have $\angle QAE=\pi/2$, $\cos\angle QAE=0$ in \eqref{bq25} and thus
\begin{align}
\left.\frac{\partial \Lambda_1}{\partial z}\cdot\frac{\partial z}{\partial \Psi}\right|_{\alpha=\alpha_{\max}}=0.
\end{align}
From Proposition 3, we have
\begin{align}
\frac{\partial\ln \Omega'}{\partial \Psi}&\approx\frac{\partial \Lambda_2}{\partial \Psi}
\end{align}
over $1\leq\alpha\leq\alpha_{\max}$. Using Proposition 5, we know that $\frac{\partial\ln \Omega'}{\partial \Psi}$ possesses a unique zero. Furthermore, from Proposition 4, since $\Psi$ is a monotonically increasing function with respect to $\alpha$, $\Omega'$ and $\Omega$ are quasi-concave with respect to $\alpha$ over $1\leq\alpha\leq\alpha_{\max}$.

From \eqref{bq6}, we know $\mathbb{E}[C_s(\alpha)]$ is a monotonically increasing function with respect to $\Omega$. According to \cite[Proposition 3.2]{Avriel} $\mathbb{E}[C_s(\alpha)]$ is a monotonically increasing function with respect to $\alpha$ over $1\leq\alpha\leq\alpha_{\max}$. $\hfill\blacksquare$

Using Proposition 2, problem \eqref{q26} can be solved by one-dimensional bisection search over $\alpha$, which yields an optimal solution.

\section{Closed-Form Near Optimal Solution for Average Secrecy Capacity Maximization}

The optimal solution obtained in Section III is a numerical solution. To reveal the boresight vector which maximizes the average secrecy capacity with the positions of the BS, the legitimate user, and the eavesdropper, we propose to investigate the secrecy capacity maximization problem considering the existence of only LoS components of wireless channels, i.e.,
\begin{equation}\label{cq1}
u_i=u_i^{\text{LoS}}
\end{equation}
for $i\in\{b,e\}$. With \eqref{cq1}, we have
\begin{equation}\label{cq2}
|h_i(\alpha)|^2=L_iG_0\Phi_i^{-1}(\alpha)
\end{equation}
and thus the secrecy capacity is
\begin{equation}\label{cq3}
C^{\text{LoS}}_s(\alpha)=\left[C^{\text{LoS}}_b(\alpha)-C^{\text{LoS}}_e(\alpha)\right]^+.
\end{equation}
where
\begin{equation}\label{cq4}
C^{\text{LoS}}_i(\alpha)=\log_2(1+\gamma L_iG_0\Phi_i^{-1}(\alpha))
\end{equation}
for $i\in\{b,e\}$. Considering the existence of only LoS components, the secrecy capacity maximization problem is formulated as
\begin{align}\label{cq5}
\max_{1\leq\alpha\leq\alpha_{\max}}\ C^{\text{LoS}}_s(\alpha).
\end{align}
The optimal solution to problem \eqref{cq5} serves as a near optimal solution to problem \eqref{q26}.

Substituting \eqref{bq3} into \eqref{cq4}, we have
\begin{equation}
C^{\text{LoS}}_i(\alpha)=\log_2\left(1+\frac{r_{i,1}\alpha+r_{i,0}}{\sqrt{d_2\alpha^2+2d_1\alpha+d_0}}\right)
\end{equation}
where
\begin{align}\label{cq7}
r_{i,0}&=\frac{\gamma L_iG_0}{\|\mathbf{q}_i\|}\mathbf{q}_e^T\mathbf{q}_i,\ \
r_{i,1}=\frac{\gamma L_iG_0}{\|\mathbf{q}_i\|}\bar{\mathbf{q}}^T\mathbf{q}_i,\\
\label{cq8}d_0&=\|\mathbf{q}_e\|^2,\ \
d_1=\bar{\mathbf{q}}^T\mathbf{q}_e,\ \
d_2=\|\bar{\mathbf{q}}\|^2.
\end{align}
Taking the first-order partial derivative of $C^{\text{LoS}}_i(\alpha)$ with respect to $\alpha$ yields
\begin{align}\label{cq15}
\frac{\partial C^{\text{LoS}}_i}{\partial \alpha}&=\frac{(r_{i,1}d_1-r_{i,0}d_2)\alpha+r_{i,1}d_0-r_{i,0}d_1}{s^2(s+r_{i,0}+r_{i,1}\alpha)\ln2}
\end{align}
where
\begin{equation}
s=\sqrt{d_2\alpha^2+2d_1\alpha+d_0}.
\end{equation}
To find the solution to the problem, we need to solve the following equation
\begin{align}\label{cq17}
\frac{\partial C^{\text{LoS}}_s}{\partial \alpha}=0.
\end{align}
Substituting \eqref{cq15} into \eqref{cq17}, we have
\begin{equation}
\frac{\rho_{b,1}\alpha+\rho_{b,0}}{s+r_{b,0}+r_{b,1}\alpha}=\frac{\rho_{e,1}\alpha+\rho_{e,0}}{s+r_{e,0}+r_{e,1}\alpha}
\end{equation}
where
\begin{align}\label{cq20}
\rho_{i,0}&=r_{i,1}d_0-r_{i,0}d_1,\\
\label{cq21}\rho_{i,1}&=r_{i,1}d_1-r_{i,0}d_2.
\end{align}
In the derivations, the property of $s>0$ is used. After some mathematical manipulations, we obtain
\begin{align}\label{cq25}
&(\rho_{b,1}-\rho_{e,1})s\alpha+(\rho_{b,0}-\rho_{e,0})s=(r_{b,1}\rho_{e,1}-r_{e,1}\rho_{b,1})\alpha^2\nonumber\\
&+(r_{b,1}\rho_{e,0}+r_{b,0}\rho_{e,1}-r_{e,1}\rho_{b,0}-r_{e,0}\rho_{b,1})\alpha\nonumber\\
&+r_{b,0}\rho_{e,0}-r_{e,0}\rho_{b,0}.
\end{align}
Because
\begin{align}
r_{b,1}\rho_{e,1}-r_{e,1}\rho_{b,1}&=d_2(r_{e,1}r_{b,0}-r_{b,1}r_{e,0}),\\
r_{b,1}\rho_{e,0}-r_{e,1}\rho_{b,0}&=d_1(r_{e,1}r_{b,0}-r_{b,1}r_{e,0}),\\
r_{b,0}\rho_{e,1}-r_{e,0}\rho_{b,1}&=d_1(r_{e,1}r_{b,0}-r_{b,1}r_{e,0}),\\
r_{b,0}\rho_{e,0}-r_{e,0}\rho_{b,0}&=d_0(r_{e,1}r_{b,0}-r_{b,1}r_{e,0}),
\end{align}
the equation \eqref{cq25} is simplified as
\begin{equation}\label{cq30}
(\rho_{b,1}-\rho_{e,1})\alpha+(\rho_{b,0}-\rho_{e,0})=s(r_{e,1}r_{b,0}-r_{b,1}r_{e,0}).
\end{equation}
Substituting \eqref{cq7} and \eqref{cq8} into \eqref{cq21}, we have
\begin{equation}
\rho_{i,1}=\frac{\gamma L_iG_0}{\|\mathbf{q}_i\|}\left(\bar{\mathbf{q}}^T\mathbf{q}_i\bar{\mathbf{q}}^T\mathbf{q}_e -\mathbf{q}_e^T\mathbf{q}_i\|\bar{\mathbf{q}}\|^2\right).
\end{equation}
Because of \eqref{bq18}, we know
\begin{equation}
\rho_{i,1}=\frac{\gamma L_iG_0w}{\|\mathbf{q}_i\|}
\end{equation}
where
\begin{equation}\label{cq32}
w=\left(\mathbf{q}_b^T\mathbf{q}_e\right)^2-\|\mathbf{q}_b\|^2\|\mathbf{q}_e\|^2.
\end{equation}
Thus, we obtain
\begin{align}
\rho_{b,1}-\rho_{e,1}=&\gamma G_0w\left(\frac{L_b}{\|\mathbf{q}_b\|}-\frac{ L_e}{\|\mathbf{q}_e\|}\right).
\end{align}
Similarly, we obtain
\begin{align}
\rho_{i,0}=\frac{\gamma L_iG_0}{\|\mathbf{q}_i\|}\left(\bar{\mathbf{q}}^T\mathbf{q}_i\|\mathbf{q}_e\|^2-\mathbf{q}_e^T\mathbf{q}_i\bar{\mathbf{q}}^T\mathbf{q}_e\right)
\end{align}
and
\begin{align}
\rho_{b,0}-\rho_{e,0}=&\frac{\gamma L_b G_0}{\|\mathbf{q}_b\|}\left(\bar{\mathbf{q}}^T\mathbf{q}_b\|\mathbf{q}_e\|^2-\mathbf{q}_e^T\mathbf{q}_b\bar{\mathbf{q}}^T\mathbf{q}_e\right)\nonumber\\
=&-\frac{\gamma L_b G_0w}{\|\mathbf{q}_b\|}.
\end{align}
Substituting \eqref{cq7} and \eqref{cq8} into \eqref{cq30}, we obtain
\begin{equation}
r_{e,1}r_{b,0}-r_{b,1}r_{e,0}=\frac{\gamma^2 L_b L_e G_0^2w}{\|\mathbf{q}_b\|\|\mathbf{q}_e\|}.
\end{equation}
It is noted that both sides of \eqref{cq30} contains the common term $w=(\mathbf{q}_b^T\mathbf{q}_e)^2-\|\mathbf{q}_b\|^2\|\mathbf{q}_e\|^2$. If $\angle BAE=0$, we know
\begin{equation}
w=\left(\mathbf{q}_b^T\mathbf{q}_e\right)^2-\|\mathbf{q}_b\|^2\|\mathbf{q}_e\|^2=0
\end{equation}
and thus $\Phi_b^{-1}(\alpha)=\Phi_e^{-1}(\alpha)$. Under this condition, according to \eqref{cq3}, we have
\begin{equation}
C^{\text{LoS}}_s(\alpha)=\left[\log_2\frac{1+\gamma L_bG_0\Phi_b^{-1}(\alpha)}{1+\gamma L_eG_0\Phi_b^{-1}(\alpha)}\right]^+.
\end{equation}
When $L_b\leq L_e$, we have $C^{\text{LoS}}_s(\alpha)=0$. When $L_b> L_e$, we have
\begin{align}
C^{\text{LoS}}_s(\alpha)&=\log_2\left(1+\frac{\gamma (L_b-L_e)G_0\Phi_b^{-1}(\alpha)}{1+\gamma L_eG_0\Phi_b^{-1}(\alpha)}\right)\nonumber\\
&=\log_2\left(1+\frac{\gamma (L_b-L_e)G_0}{\Phi_b(\alpha)+\gamma L_eG_0}\right)
\end{align}
which is a monotonically decreasing function with respect to $\alpha$. The optimal $\alpha$ is $\alpha=1$.

If $\angle BAE\neq0$, since $\gamma G_0>0$, we have
\begin{align}\label{cq41}
&\left(\frac{L_b}{\|\mathbf{q}_b\|}-\frac{ L_e}{\|\mathbf{q}_e\|}\right)\alpha-\frac{L_b}{\|\mathbf{q}_b\|}\nonumber\\
&=\frac{\gamma L_b L_e G_0}{\|\mathbf{q}_b\|\|\mathbf{q}_e\|}\sqrt{\|\bar{\mathbf{q}}\|^2\alpha^2+2\bar{\mathbf{q}}^T\mathbf{q}_e \alpha+\|\mathbf{q}_e\|^2}.
\end{align}

When $L_b/\|\mathbf{q}_b\|< L_e/\|\mathbf{q}_e\|$, from \eqref{cq15} to \eqref{cq41}, we know
\begin{align}\label{cq42}
\frac{\partial C^{\text{LoS}}_s(\alpha)}{\partial \alpha}=\frac{\gamma G_0 w\left(\Upsilon_1-\Upsilon_2\right)}{s(\ln2)\prod_{i\in\{b,e\}}(s+r_{i,0}+r_{i,1}\alpha)}
\end{align}
where
\begin{align}
\Upsilon_1&=\left(\frac{L_b}{\|\mathbf{q}_b\|}-\frac{ L_e}{\|\mathbf{q}_e\|}\right)\alpha-\frac{L_b}{\|\mathbf{q}_b\|},\\
\label{aq103}\Upsilon_2&=\frac{\gamma L_b L_e G_0}{\|\mathbf{q}_b\|\|\mathbf{q}_e\|}\sqrt{\|\bar{\mathbf{q}}\|^2\alpha^2+2\bar{\mathbf{q}}^T\mathbf{q}_e \alpha+\|\mathbf{q}_e\|^2}.
\end{align}
It is noted that $\Upsilon_1$ and $\Upsilon_2$ are the left-hand and right-hand sides of \eqref{cq41}, respectively. In \eqref{cq42}, we have
\begin{align}
r_{i,0}+r_{i,1}\alpha&=\frac{\gamma L_iG_0}{\|\mathbf{q}_i\|}(\mathbf{q}_e^T\mathbf{q}_i+\bar{\mathbf{q}}^T\mathbf{q}_i\alpha)\nonumber\\
&=\frac{\gamma L_iG_0}{\|\mathbf{q}_i\|}\mathbf{q}_i^T\mathbf{q}_a(\bm{\theta}).
\end{align}
Because of the constraint $1\leq\alpha\leq\alpha_{\max}$, we know
\begin{align}
r_{i,0}+r_{i,1}\alpha\geq0.
\end{align}
Since $w<0$, the sign of $\frac{\partial C^{\text{LoS}}_s(\alpha)}{\partial \alpha}$ is determined by the term $\Upsilon_1-\Upsilon_2$. If
\begin{align}
\frac{L_b}{\|\mathbf{q}_b\|}<\frac{ L_e}{\|\mathbf{q}_e\|}
\end{align}
and thus $\Upsilon_1-\Upsilon_2<0$, we have 
\begin{equation}
\frac{\partial C^{\text{LoS}}_s(\alpha)}{\partial \alpha}>0.
\end{equation}
The optimal $\alpha$ is $\alpha^o=\alpha_{\max}$.

In the following, we consider
\begin{align}\label{cq43}
\frac{L_b}{\|\mathbf{q}_b\|}\geq\frac{ L_e}{\|\mathbf{q}_e\|}.
\end{align}
Squaring both sides of \eqref{cq41} yields
\begin{equation}\label{cq45}
\zeta_2\alpha^2+\zeta_1\alpha+\zeta_0=0
\end{equation}
where
\begin{align}
\zeta_2&=(L_b\|\mathbf{q}_e\|-L_e\|\mathbf{q}_b\|)^2-\kappa\|\bar{\mathbf{q}}\|^2,\\
\zeta_1&=-2L_b^2\|\mathbf{q}_e\|^2+2L_bL_e\|\mathbf{q}_b\|\|\mathbf{q}_e\|-2\kappa\bar{\mathbf{q}}^T\mathbf{q}_e,\\
\zeta_0&=(L_b^2-\kappa)\|\mathbf{q}_e\|^2,\\
\label{cq51}\kappa&=(\gamma L_b L_e G_0)^2.
\end{align}
After some mathematical manipulations, the discriminant of the quadratic equation \eqref{cq45} is derived as follows
\begin{align}\label{cq52}
\zeta_1^2-4\zeta_2\zeta_0&=4\kappa^2w+4\kappa\|(L_b\|\mathbf{q}_b\|\mathbf{q}_e-L_e\|\mathbf{q}_e\|\mathbf{q}_b)\|^2.
\end{align}
In \eqref{cq52}, since $\angle BAE\neq0$ and thus $w<0$, the sign of $\zeta_1^2-4\zeta_2\zeta_0$ is undetermined. We consider three cases of $\zeta_1^2-4\zeta_2\zeta_0<0$, $\zeta_1^2-4\zeta_2\zeta_0=0$, and $\zeta_1^2-4\zeta_2\zeta_0>0$.

The case of $\zeta_1^2-4\zeta_2\zeta_0<0$ is satisfied when $\kappa$ is sufficiently large. This is because with the increase of $\kappa$, $\kappa^2w$ decreases whereas  $\kappa\|(L_b\|\mathbf{q}_b\|\mathbf{q}_e-L_e\|\mathbf{q}_e\|\mathbf{q}_b)\|^2$ increases. The former decreases at a much higher rate than the latter increases. From \eqref{cq51}, we know that $\kappa$ is a monotonically increasing function with respect to $\gamma$. Denote $\gamma_0$ as the solution to equation 
\begin{align}
\kappa w+\|(L_b\|\mathbf{q}_b\|\mathbf{q}_e-L_e\|\mathbf{q}_e\|\mathbf{q}_b)\|^2=0.
\end{align}
If $\gamma>\gamma_0$, we have $\zeta_1^2-4\zeta_2\zeta_0<0$. Under this condition, it can be verified that
\begin{equation}\label{cq53}
\alpha^o=\alpha_{\max}.
\end{equation}
Therefore, we have $\mathbf{q}^T_a(\bm{\theta})\mathbf{q}_e=0$, which indicates that the optimal boresight vector is orthogonal to the vector $\mathbf{q}_e$.

The case of $\zeta_1^2-4\zeta_2\zeta_0=0$ is satisfied when $\gamma=\gamma_0$ or $\gamma\rightarrow0$. Under the former condition, the optimal $\alpha$ is also $\alpha_{\max}$. Under the latter condition, the optimal $\alpha$ is
\begin{equation}
\alpha^o=-\frac{\zeta_1}{2\zeta_2}=\frac{L_b\|\mathbf{q}_e\|}{L_b\|\mathbf{q}_e\|-L_e\|\mathbf{q}_b\|}=\left(1-\frac{L_e\|\mathbf{q}_b\|}{L_b\|\mathbf{q}_e\|}\right)^{-1}.
\end{equation}

The case of $\zeta_1^2-4\zeta_2\zeta_0>0$ is satisfied when $0<\gamma<\gamma_0$. Under this condition, two roots of equation \eqref{cq45} are
\begin{equation}
\alpha=\frac{-\zeta_1\pm\sqrt{\zeta_1^2-4\zeta_2\zeta_0}}{2\zeta_2}.
\end{equation}
To proceed, we need the following proposition.

\emph{Proposition 7}: Consider the case of $\zeta_1^2-4\zeta_2\zeta_0>0$ and the condition of $L_b/\|\mathbf{q}_b\|\geq L_e/\|\mathbf{q}_e\|$. When $\zeta_2>0$, the optimal $\alpha$ is 
\begin{equation}\label{cq62}
\alpha^o=\frac{-\zeta_1+\sqrt{\zeta_1^2-4\zeta_2\zeta_0}}{2\zeta_2}.
\end{equation}

When $\zeta_2<0$ and $\tilde{\alpha}<\alpha^\star$ where
\begin{align}\label{cq63}
\tilde{\alpha}=\left(1-\frac{L_e\|\mathbf{q}_b\|}{L_b\|\mathbf{q}_e\|}\right)^{-1} \text{ and }
\alpha^\star=-\frac{\bar{\mathbf{q}}^T\mathbf{q}_e}{\|\bar{\mathbf{q}}\|^2},
\end{align}
the optimal $\alpha$ is also \eqref{cq62}. When $\zeta_2<0$ and $\tilde{\alpha}\geq\alpha^\star$, the optimal $\alpha$ is $\alpha^o=\alpha_{\max}$.

When $\zeta_2=0$ and $\tilde{\alpha}<\alpha^\star$, the optimal $\alpha$ is
\begin{align}\label{cq65}
\alpha^o=\frac{(L_b^2-\kappa)\|\mathbf{q}_e\|^2}{2L_b^2\|\mathbf{q}_e\|^2-2L_bL_e\|\mathbf{q}_b\|\|\mathbf{q}_e\|+2\kappa\bar{\mathbf{q}}^T\mathbf{q}_e}.
\end{align}
When $\zeta_2=0$ and $\tilde{\alpha}\geq\alpha^\star$, the optimal $\alpha$ is $\alpha^o=\alpha_{\max}$.

\emph{Proof}: See Appendix E. $\hfill\blacksquare$

\section{Secrecy Outage Probability At High SNR}

At high SNR, Section IV provides us a closed-form optimal solution. Using the expression of optimal solution, we are able to theoretically analyze the system secrecy outage probability.

If the SNR is sufficiently high, i.e., $\gamma>\gamma_0$, substituting the closed-form optimal solution \eqref{cq53} into \eqref{q7}, we know $h_e(\bm{\theta})=0$ and thus $C_e=0$. The system secrecy outage probability is given by
\begin{equation}\label{dq1}
\text{SOP}(R_s)=\text{Pr}(C_s<R_s)=\text{Pr}(C_b<R_s)
\end{equation}
where $R_s$ denotes a predefined secrecy rate at which the BS transmits confidential information to the legitimate user.

Substituting \eqref{q12} into \eqref{dq1}, we have
\begin{equation}
\text{SOP}(R_s)=\text{Pr}(\left|h_b(\alpha_{\max})\right|^2<\gamma_{\text{th}})
\end{equation}
where
\begin{equation}
\gamma_{\text{th}}=\gamma^{-1}(2^{R_s}-1).
\end{equation}
According to \cite[Eqn. (9)]{Bhargav}, we have
\begin{equation}\label{dq4}
\text{SOP}(R_s)=1-Q_1\left(2\sqrt{K_b},\sqrt{\frac{2(1+K_b)\gamma_{\text{th}}}{\mathbb{E}[|h_b(\alpha_{\max})|^2]}}\right)
\end{equation}
where $Q_1(x,y)=\int_{y}^{\infty}t\exp(-t^2/2-x^2/2)I_0(xt)dt$ is the first-order Marcum $Q$-function. From \eqref{bq2}, we know 
\begin{equation}\label{dq5}
\mathbb{E}[|h_b(\alpha_{\max})|^2]=L_bG_0\Phi_b^{-1}(\alpha_{\max})
\end{equation}
where
\begin{align}
\Phi_b^{-1}(\alpha_{\max})=\frac{(\mathbf{q}_e+\alpha_{\max}(\mathbf{q}_b-\mathbf{q}_e))^T\mathbf{q}_b}{\|\mathbf{q}_e+\alpha_{\max}(\mathbf{q}_b-\mathbf{q}_e)\| \|\mathbf{q}_b\|}.
\end{align}
From \eqref{q24}, we know $\mathbf{q}_e+\alpha_{\max}(\mathbf{q}_b-\mathbf{q}_e)$ is orthogonal to $\mathbf{q}_e$ which shows that $\angle QAE=\pi/2$ achieves optimality. Thus, we have
\begin{align}\label{dq7}
\Phi_b^{-1}(\alpha_{\max})=\cos \angle QAB=\sin\angle BAE.
\end{align}
Substituting \eqref{dq7} and \eqref{dq5} into \eqref{dq4}, the system secrecy outage probability at high SNR is
\begin{equation}
\text{SOP}(R_s)=1-Q_1\left(2\sqrt{K_b},\sqrt{\frac{2(1+K_b)\gamma_{\text{th}}}{L_bG_0\sin\angle BAE}}\right).
\end{equation}

\section{Simulation Results}

In simulations, the rotatable antenna-assisted secure communication system operates at a carrier frequency of 2.4 GHz, corresponding to a wavelength of $\lambda = 0.125$ m \cite{LDai25}. The channel power gain at the reference distance of one meter is $\zeta_0=0.001$. The path loss exponents are $\beta_b=\beta_e=3$ \cite{LDai25}. The noise power is $\sigma^2=-60$ dBm \cite{LDai25}. Without loss of generality, we assume that both the legitimate user and eavesdropper are located on the $x-z$ plane. If not specified, the distances from the BS to the legitimate user and  eavesdropper are 50 m and 70 m, respectively. The Rician factors are $K_b=K_e=1$. The angles between the $x$-axis and the links from BS to legitimate user and eavesdropper are $60^\circ$ and $30^\circ$, respectively. Therefore, the position vectors of the legitimate user and eavesdropper are 
\begin{align}
\mathbf{q}_b = [50\cos(60^\circ), 0, 50\sin(60^\circ)]^T,\\
\mathbf{q}_e = [70\cos(30^\circ), 0, 70\sin(30^\circ)]^T,
\end{align}
respectively. Under this condition, $\alpha_{\max}$ in \eqref{q25} is $2.62$.

In Fig. 2, we present the average secrecy capacity $\mathbb{E}[C_s(\alpha)]$ and the secrecy capacity $C^{\text{LoS}}_s(\alpha)$ versus $\alpha$ to verify our proposed optimal and near optimal solutions, where the transmission power of the BS is $P=16$ dBm.  From Fig. 2, it is shown that the average secrecy capacity is a quasi-concave function with respect to $\alpha$. Furthermore, the optimum which maximizes $C^{\text{LoS}}_s(\alpha)$ instead of the average secrecy capacity provides us the closed-form near optimal solution. From Fig. 2, compared with the optimum of the average secrecy capacity, it is found that the near optimal solution has only about 0.003 bps/Hz performance loss.  

\begin{figure}
\centering
\includegraphics[width=3.6in]{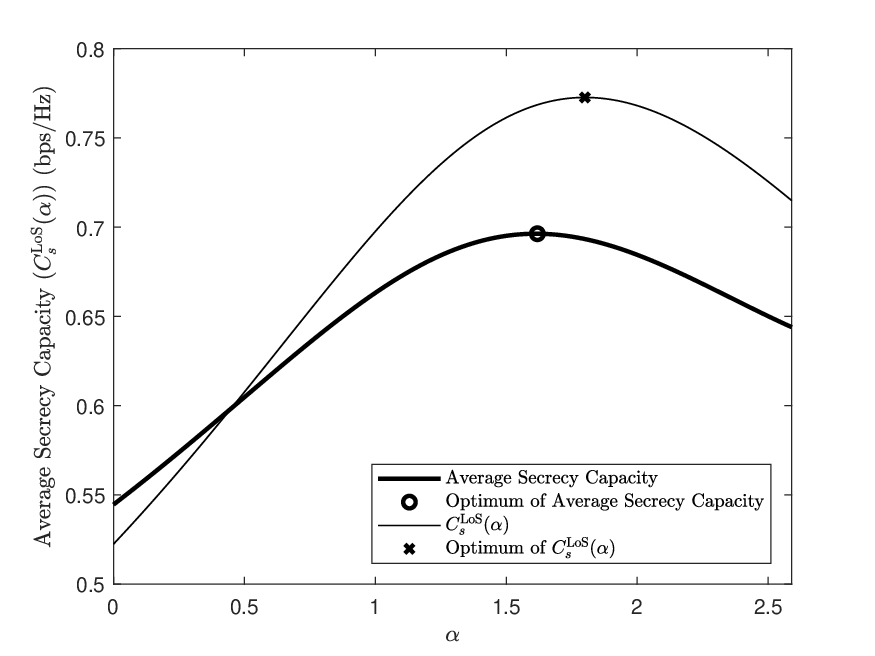}
\caption{Average secrecy capacity $\mathbb{E}[C_s(\alpha)]$ and the secrecy capacity $C^{\text{LoS}}_s(\alpha)$ versus $\alpha$; performance comparison of our proposed optimal and near optimal solutions, where $P=16$ dBm.}
\end{figure}

In Fig. 3, we present the average secrecy capacity achieved by our proposed optimal and near optimal solutions, denoted as ``Optimal" and ``Near Optimal" in the legend, respectively, for different values of $K_b=K_e=K$. The position of eavesdropper is $\mathbf{q}_e = [70\cos \upsilon, 0, 70\sin \upsilon]^T$, where $\upsilon$ is $0^\circ$ and $30^\circ$. From Fig. 3, it is observed that the near optimal solution achieves almost the same average secrecy capacity as the optimal solution for different values of $P$. Furthermore, the system with $\upsilon=0^\circ$ performs better than that with $\upsilon=30^\circ$. This shows the importance of the relative positions of the BS, the legitimate user and eavesdropper. From Fig. 3, it is also found that the system with $K=5$ performs better than that with $K=1$.

\begin{figure}
\centering
\includegraphics[width=3.6in]{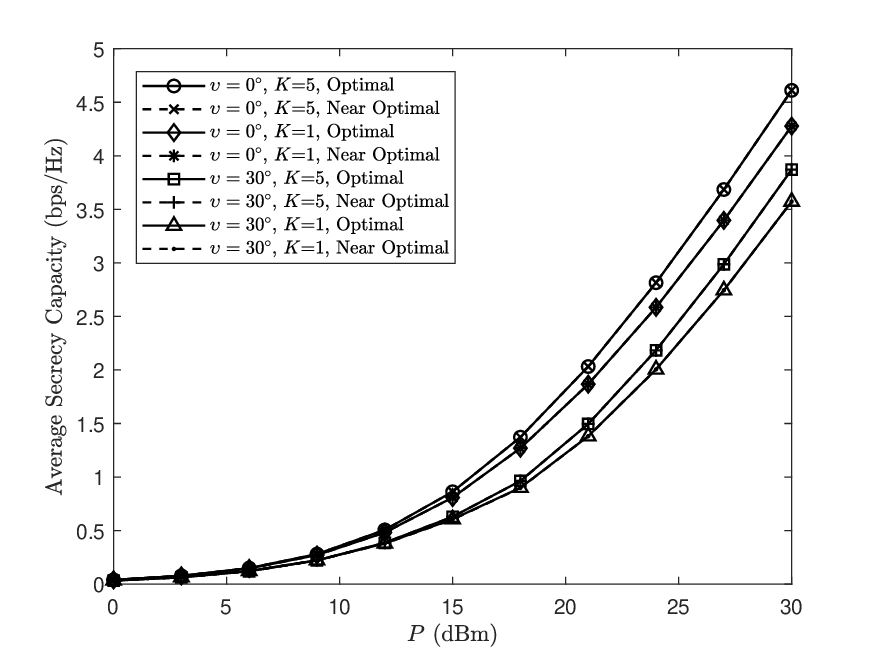}
\caption{Average secrecy capacity $\mathbb{E}[C_s(\alpha)]$ versus $P$; performance comparison of our proposed optimal and near optimal solutions.}
\end{figure}

In Fig. 4, we present the simulation and theoretical secrecy outage probabilities for different values of predefined secrecy rate $R_s$, where the transmission power of the BS is $P=25$ dBm. The position of eavesdropper is $\mathbf{q}_e = [70\cos \upsilon, 0, 70\sin \upsilon]^T$, where $\upsilon$ is $0^\circ$, $30^\circ$, and $45^\circ$. From Fig. 4, it is observed that the theoretical secrecy outage probabilities match the simulation ones. This is because $P=25$ dBm is sufficiently high to achieve high SNR and thus the analytical result derived in Section V is accurate.

\begin{figure}
\centering
\includegraphics[width=3.6in]{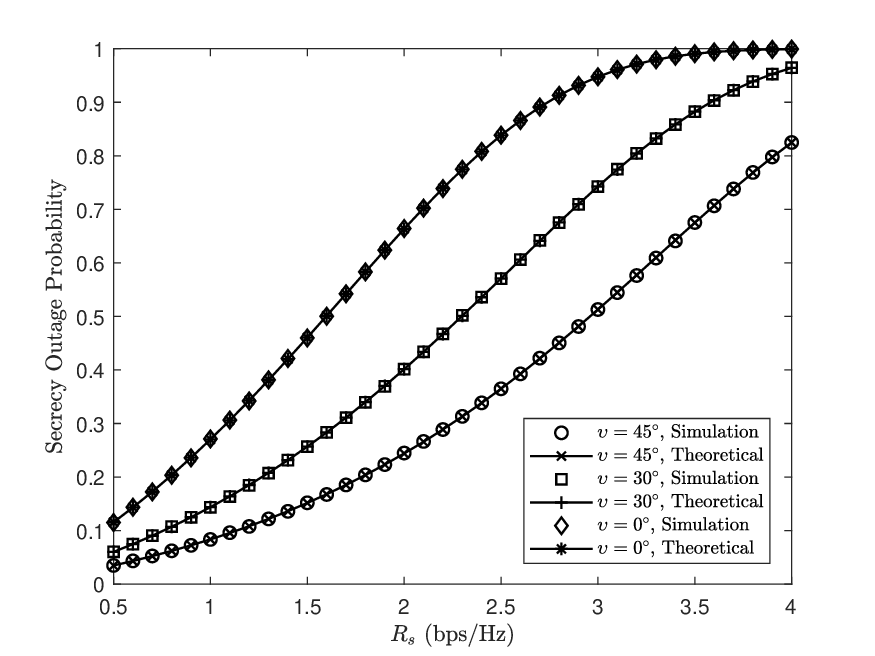}
\caption{Secrecy outage probability versus $R_s$; comparison of simulated and theoretical secrecy outage probabilities, where $P=25$ dBm.}
\end{figure}

In Fig. 5, we present the simulation and theoretical secrecy outage probabilities for different values of $P$, where $R_s=1$ bps/Hz. From Fig. 5, it is illustrated that for different values of $P$, the theoretical secrecy outage probabilities match the simulation ones when $\upsilon$ is $0^\circ$. When $\upsilon$ is $30^\circ$ and $45^\circ$, $P$ should be larger than 21 dBm and 24 dBm, respectively, to ensure the accurate analytical results derived in Section V. Otherwise, the analytical results are upper bounds on the simulation results.

\begin{figure}
\centering
\includegraphics[width=3.6in]{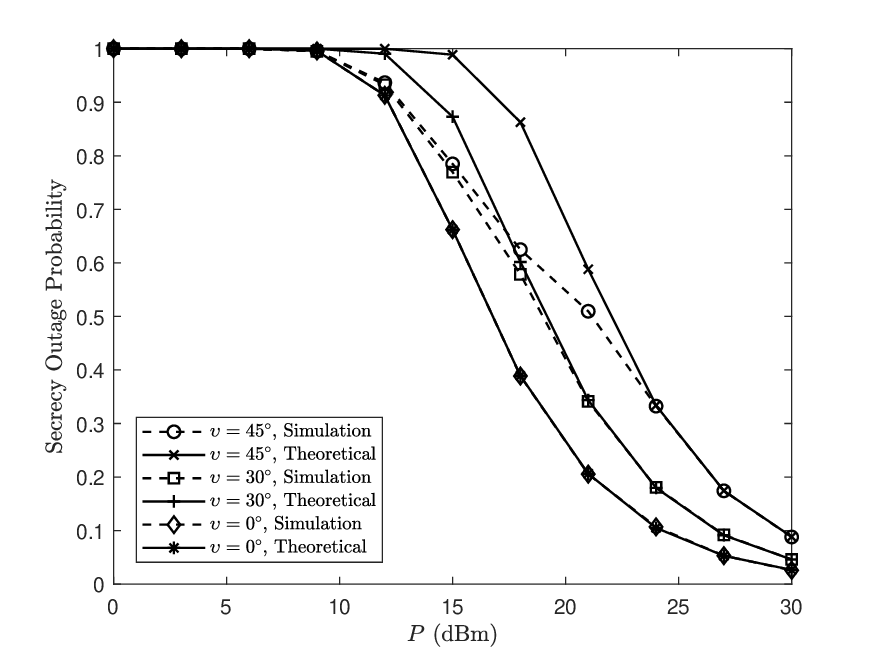}
\caption{Secrecy outage probability versus $P$; comparison of simulated and theoretical secrecy outage probabilities, where $R_s=1$ bps/Hz.}
\end{figure}

\section{Conclusion}

In this paper, we have investigated the average secrecy rate maximization problem of a rotatable antenna-assisted secure communication system, where the geometric property of optimal boresight vector and the quasi-concavity of the optimization problem are rigorously proved. Furthermore, we have derived a closed-form near optimal solution considering the existence of only LoS components of wireless channels and theoretically characterized the secrecy outage probability at high SNR. Simulation results have validated that the near optimal solution achieves almost the same average secrecy capacity as the optimal solution and at high SNR, the theoretical secrecy outage probabilities match the simulation ones.

\appendices

\section{Proof of Proposition 3}

Taking the first-order partial derivative of $\Lambda_1$ with respect to $z$ yields
\begin{align}\label{aq1}
\frac{\partial \Lambda_1}{\partial z}&=\frac{\gamma(y'-x')}{\ln\left( \frac{z+\gamma x'}{z+\gamma y'} \right)(z+\gamma x')(z+\gamma y')}.
\end{align}
To derive the lower and upper bounds on $\frac{\partial \Lambda_1}{\partial z}$, we take the second-order partial derivative of $\Lambda_1$ with respect to $z$. After some mathematical manipulations, we obtain
\begin{align}
\frac{\partial^2 \Lambda_1}{\partial z^2}&=Y_1\cdot Y_2
\end{align}
where
\begin{align}
\label{aq3}Y_1&=\frac{\gamma(x'-y')}{\ln\left(\frac{z+\gamma x'}{z+\gamma y'} \right)(z+\gamma x')^2(z+\gamma y')^2},\\
\label{aq4}Y_2&=\frac{\gamma(y'-x')}{\ln\left( \frac{z+\gamma x'}{z+\gamma y'} \right)}+ 2z + \gamma (x'+y').
\end{align}

From \eqref{bq10}, the integral interval of $y'$ is over $(0,x']$. Under this condition, we know $x'\geq y'$ and thus $ Y_1\geq0 $. To prove that $\frac{\partial \Lambda_1}{\partial z}$ is a monotonically increasing function with respect to $z$, it is required to prove $Y_2>0$.

To prove $Y_2>0$, we use the following property
\begin{equation}\label{aq5}
\ln(1+x)\geq\frac{x}{(1+x)}
\end{equation}
for $x\geq0$, which can be proved as follows. Define
\begin{equation}
\vartheta(x)=\ln(1+x)-\frac{x}{(1+x)}.
\end{equation}
We know $\vartheta(0)=0$ and
\begin{equation}
\frac{d\vartheta(x)}{d x}=\frac{x}{(1+x)^2}\geq0
\end{equation}
for $x\geq0$. Using \eqref{aq5}, we have
\begin{equation}\label{aq8}
\ln\left(\frac{z+\gamma x'}{z+\gamma y'} \right)
= \ln\left(1+ \frac{\gamma(x'-y') }{z+\gamma y'} \right)
\geq\frac{\gamma(x'-y') }{z+\gamma x'}.
\end{equation}
Substituting \eqref{aq8} into \eqref{aq4}, we obtain
\begin{align}
Y_2\geq z + \gamma y'\geq0.
\end{align}
Therefore, $\frac{\partial \Lambda_1}{\partial z}$ is a monotonically increasing function with respect to $z$.

In the following, we derive the lower and upper bounds on $\frac{\partial \Lambda_1}{\partial z}$. Because $z \geq 1$, from \eqref{aq1}, we know
\begin{align}
\frac{\partial \Lambda_1}{\partial z}\geq \Theta
\end{align}
where
\begin{align}
\Theta=\frac{\gamma(y'-x')}{\ln\left( \frac{1+\gamma x'}{1+\gamma y'} \right)(1+\gamma x')(1+\gamma y')}.
\end{align}
The expression $\Theta$ contains unknown parameters $x'$ and $y'$ with $x'\geq y' \geq 0$. Taking the first-order partial derivative of $\Theta$ with respect
to $x'$, after some mathematical manipulations, we have
\begin{align}
\frac{\partial \Theta}{\partial x'} = -\frac{\gamma\xi}{\ln\left( \frac{1+\gamma x'}{1+\gamma y'}\right)(1+\gamma x')^2}
\end{align}
where
\begin{align}\label{aq13}
\xi=\frac{\gamma(y'-x')}{(1+\gamma y')\ln\left( \frac{1+\gamma x'}{1+\gamma y'}\right)}+1.
\end{align}
From the first-order Taylor expansions of $\ln(1+x)$ about $x=0$, we know
\begin{equation}\label{aq14}
\ln(1+x)\leq x
\end{equation}
because $\ln(1+x)$ is a concave function with respect to $x$. Employing \eqref{aq14}, we have
\begin{equation}\label{aq15}
\ln\left(\frac{1+\gamma x'}{1+\gamma y'} \right)
= \ln\left(1+ \frac{\gamma(x'-y') }{1+\gamma y'} \right)
\leq\frac{\gamma(x'-y') }{1+\gamma y'}.
\end{equation}
Substituting \eqref{aq15} into \eqref{aq13}, we obtain
$\xi\leq 0$ and thus
\begin{align}
\frac{\partial \Theta}{\partial x'}\geq0,
\end{align}
which shows $\Theta$ is a monotonically increasing function with respect to $x'\geq0$. The minimum value of $\Theta$ is achieved when $x'\rightarrow y'$ and $y'\rightarrow 0$ since $x'\geq y' \geq 0$.

Using L'Hospital's rule, we have
\begin{equation}
\lim_{y' \rightarrow 0}\lim_{x' \rightarrow y'}\Theta=\lim_{y' \rightarrow 0}-\frac{1}{1+\gamma y'}=-1.
\end{equation}

The upper bound on $\frac{\partial \Lambda_1}{\partial z}$ is achieved when $z \rightarrow \infty$,
\begin{equation}
\lim_{z \rightarrow \infty}\frac{\partial \Lambda_1}{\partial z}=\lim_{z \rightarrow \infty}\frac{\varpi_1}{\varpi_2}
\end{equation}
where
\begin{align}
\varpi_1=&\frac{\gamma(y'-x')}{(z+\gamma x')(z+\gamma y')},\\
\varpi_2=&\ln\left( \frac{z+\gamma x'}{z+\gamma y'} \right).
\end{align}
Because $\lim_{z \rightarrow \infty}\varpi_1=0$ and $\lim_{z \rightarrow \infty}\varpi_2=0$, to apply L'Hospital's rule, we obtain
\begin{align}
\frac{\partial\varpi_1}{\partial z}=&-\frac{\gamma(y'-x')(2z+\gamma(x'+y'))}{(z+\gamma x')^2(z+\gamma y')^2},\\
\frac{\partial\varpi_2}{\partial z}=&\frac{\gamma(y'-x')}{(z+\gamma x')(z+\gamma y')}.
\end{align}
Applying L'Hospital's rule yields
\begin{equation}
\lim_{z \rightarrow \infty}\frac{\partial \Lambda_1}{\partial z}=\lim_{z \rightarrow \infty} -\frac{2z+\gamma(x'+y')}{(z+\gamma x')(z+\gamma y')} = 0.
\end{equation}
This completes the proof.

\section{Proof of Proposition 4}

In \eqref{bq19}, the sign of $\frac{\partial\Psi}{\partial \alpha}$ is determined by the $\bar{\mathbf{q}}^T\mathbf{q}_b\|\mathbf{q}_e\|^2-\bar{\mathbf{q}}^T\mathbf{q}_e\mathbf{q}_e^T\mathbf{q}_b$.
Because
\begin{align}\label{aq18}
\frac{\bar{\mathbf{q}}^T\mathbf{q}_b}{\|\bar{\mathbf{q}}\|\|\mathbf{q}_b\|}&=\cos\angle ABE,\\
\label{aq19}\frac{\bar{\mathbf{q}}^T\mathbf{q}_e}{\|\bar{\mathbf{q}}\|\|\mathbf{q}_e\|}&=\cos (\pi-\angle AEB),\\
\label{aq20}\frac{\mathbf{q}_e^T\mathbf{q}_b}{\|\mathbf{q}_e\|\|\mathbf{q}_b\|}&=\cos\angle BAE,
\end{align}
we have
\begin{align}
&\frac{\bar{\mathbf{q}}^T\mathbf{q}_b\|\mathbf{q}_e\|^2-\bar{\mathbf{q}}^T\mathbf{q}_e\mathbf{q}_e^T\mathbf{q}_b}{\|\bar{\mathbf{q}}\|\|\mathbf{q}_b\|\|\mathbf{q}_e\|^2}\nonumber\\
=&\cos\angle ABE-\cos (\pi-\angle AEB)\cos\angle BAE.
\end{align}
Since
\begin{equation}
\angle ABE=(\pi-\angle AEB)-\angle BAE,
\end{equation}
according to cosine of difference formula, we obtain
\begin{align}
&\frac{\bar{\mathbf{q}}^T\mathbf{q}_b\|\mathbf{q}_e\|^2-\bar{\mathbf{q}}^T\mathbf{q}_e\mathbf{q}_e^T\mathbf{q}_b}{\|\bar{\mathbf{q}}\|\|\mathbf{q}_b\|\|\mathbf{q}_e\|^2}\nonumber\\
=&\sin (\pi-\angle AEB)\sin\angle BAE\geq 0.
\end{align}
Therefore, we know that $\Psi$ is a monotonically increasing function with respect to $\alpha$.

\section{Proof of Proposition 5}

By defining $\tau=\sqrt{K_e\eta_e \Psi y'}\geq 0$, we rewrite \eqref{bq20} as
\begin{equation}
\frac{\partial \Lambda_2}{\partial \Psi}=\frac{1}{\Psi}\left(1+\tau\mathcal{I}(\tau)\right)-\eta_ey'
\end{equation}
where
\begin{align}
\mathcal{I}(\tau)=\frac{I_1(2\tau)}{I_0(2\tau)}.
\end{align}
Taking the second-order partial derivative of $\Lambda_2$ with respect to $\Psi$ yields
\begin{align}\label{aq25}
\frac{\partial^2 \Lambda_2}{\partial \Psi^2}&=\frac{1}{\Psi}\left(\mathcal{I}(\tau)+\tau\frac{\partial\mathcal{I}(\tau)}{\partial\tau}\right)\frac{\partial \tau}{\partial \Psi}-\frac{1}{\Psi^2}\left(1+\tau\mathcal{I}(\tau)\right)
\end{align}
where
\begin{align}
\frac{\partial\tau}{\partial \Psi}&=\frac{K_e\eta_e y'}{2\sqrt{K_e\eta_e \Psi y'}}=\frac{\tau}{2\Psi},\\
\label{aq27}\frac{\partial\mathcal{I}(\tau)}{\partial\tau}&=2(1-\mathcal{I}^2(\tau))-\mathcal{I}(\tau)/\tau.
\end{align}
In the derivations, the property $I_1'(x)=I_0(x)-I_1(x)/x$ is employed \cite[Eqn. 9.6.26]{Abramowitz}. After some mathematical manipulations, we re-express \eqref{aq25} as
\begin{align}
\frac{\partial^2 \Lambda_2}{\partial \Psi^2}&=\frac{1}{\Psi^2}\Delta
\end{align}
where
\begin{align}
\Delta=\tau^2-\tau^2\mathcal{I}^2(\tau)-\tau\mathcal{I}(\tau)-1.
\end{align}
Taking the first-order partial derivative of $\Delta$ with respect to $\tau$ yields
\begin{align}\label{aq30}
\frac{\partial\Delta}{\partial\tau}=2\tau-2\tau\mathcal{I}^2(\tau)-2\tau^2\mathcal{I}(\tau)\frac{\partial\mathcal{I}(\tau)}{\partial\tau}-\mathcal{I}(\tau)-\tau\frac{\partial\mathcal{I}(\tau)}{\partial\tau}.
\end{align}
Using \eqref{aq27}, we simplify \eqref{aq30} as
\begin{align}
\frac{\partial\Delta}{\partial\tau}=-2\tau^2\mathcal{I}(\tau)\frac{\partial\mathcal{I}(\tau)}{\partial\tau}.
\end{align}
From \cite[Theorem 1]{Segura}, $\mathcal{I}(\tau)$ is a monotonically increasing function for $\tau\geq0$, we have
\begin{align}
\frac{\partial\mathcal{I}(\tau)}{\partial\tau}\geq0 \text{ and }\frac{\partial\Delta}{\partial\tau}\leq0.
\end{align}
Since
\begin{align}
\Delta|_{\tau=0}=-1,
\end{align}
we know
\begin{align}
\Delta\leq0  \text{ and }\frac{\partial^2 \Lambda_2}{\partial \Psi^2}\leq0.
\end{align}
Furthermore, from \eqref{bq20} and \eqref{bq21}, Proposition 5 is proved.

\section{Proof of Proposition 6}

Taking the first-order partial derivative of $z$ with respect to $\alpha$ yields
\begin{align}\label{aq41}
&\frac{\partial z}{\partial \alpha}=\frac{\partial \Phi_b(\alpha)}{\partial \alpha}=\|\mathbf{q}_b\|\\
&\cdot\frac{(\|\bar{\mathbf{q}}\|^2\mathbf{q}_e^T\mathbf{q}_b-\mathbf{q}_e^T\bar{\mathbf{q}}\bar{\mathbf{q}}^T\mathbf{q}_b)\alpha+\mathbf{q}_e^T\mathbf{q}_b\mathbf{q}_e^T\bar{\mathbf{q}}-\|\mathbf{q}_e\|^2\bar{\mathbf{q}}^T\mathbf{q}_b}{\left(\mathbf{q}_e^T\mathbf{q}_b+\alpha\bar{\mathbf{q}}^T\mathbf{q}_b\right)^2\|\mathbf{q}_e+\alpha\bar{\mathbf{q}}\| }.\nonumber
\end{align}
Taking the first-order partial derivative of $\Psi$ with respect to $\alpha$ yields
\begin{align}\label{aq42}
\frac{\partial \Psi}{\partial \alpha}&=\frac{\|\mathbf{q}_e\|}{\|\mathbf{q}_b\|}\frac{\partial (\mathbf{q}_e^T\mathbf{q}_b+\alpha\bar{\mathbf{q}}^T\mathbf{q}_b)/(\|\mathbf{q}_e\|^2+\alpha\mathbf{q}_e^T\bar{\mathbf{q}})}{\partial \alpha}\\
&=\frac{\|\mathbf{q}_e\|}{\|\mathbf{q}_b\|}\cdot\frac{\|\mathbf{q}_e\|^2\bar{\mathbf{q}}^T\mathbf{q}_b-\mathbf{q}_e^T\mathbf{q}_b\mathbf{q}_e^T\bar{\mathbf{q}}}{(\|\mathbf{q}_e\|^2+\alpha\mathbf{q}_e^T\bar{\mathbf{q}})^2}.
\end{align}
Using \eqref{aq41} and \eqref{aq42}, we obtain
\begin{equation}
\frac{\partial z}{\partial \Psi}=\frac{\frac{\partial z}{\partial \alpha}}{\frac{\partial \Psi}{\partial \alpha}}=\chi\left(\frac{\|\bar{\mathbf{q}}\|^2\mathbf{q}_e^T\mathbf{q}_b-\mathbf{q}_e^T\bar{\mathbf{q}}\bar{\mathbf{q}}^T\mathbf{q}_b}{\|\mathbf{q}_e\|^2\bar{\mathbf{q}}^T\mathbf{q}_b-\mathbf{q}_e^T\mathbf{q}_b\mathbf{q}_e^T\bar{\mathbf{q}}}\alpha-1\right)
\end{equation}
where
\begin{equation}\label{aq49}
\chi=\frac{\|\mathbf{q}_b\|^2}{\|\mathbf{q}_e\|}\cdot\frac{(\|\mathbf{q}_e\|^2+\alpha\mathbf{q}_e^T\bar{\mathbf{q}})^2}{\left(\mathbf{q}_e^T\mathbf{q}_b+\alpha\bar{\mathbf{q}}^T\mathbf{q}_b\right)^2\|\mathbf{q}_e+\alpha\bar{\mathbf{q}}\|}>0.
\end{equation}
From \eqref{aq18}, \eqref{aq19}, \eqref{aq20}, we know
\begin{align}\label{aq50}
\|\bar{\mathbf{q}}\|^2\mathbf{q}_e^T\mathbf{q}_b &=\|\bar{\mathbf{q}}\|^2\|\mathbf{q}_e\|\|\mathbf{q}_b\|\cos\angle BAE,\\
\label{aq51}\mathbf{q}_e^T\bar{\mathbf{q}}\bar{\mathbf{q}}^T\mathbf{q}_b&=\|\bar{\mathbf{q}}\|^2\|\mathbf{q}_e\|\|\mathbf{q}_b\|\cos(\pi-\angle AEB)\cos\angle ABE,\\
\label{aq52}\|\mathbf{q}_e\|^2\bar{\mathbf{q}}^T\mathbf{q}_b&=\|\mathbf{q}_e\|^2\|\bar{\mathbf{q}}\|\|\mathbf{q}_b\|\cos\angle ABE,\\
\label{aq53}\mathbf{q}_e^T\mathbf{q}_b\mathbf{q}_e^T\bar{\mathbf{q}}&=\|\mathbf{q}_e\|^2\|\bar{\mathbf{q}}\|\|\mathbf{q}_b\|\cos\angle BAE\cos(\pi-\angle AEB).
\end{align}
Furthermore, we have
\begin{equation}\label{aq54}
\pi-\angle AEB=\angle BAE+\angle ABE.
\end{equation}
Using \eqref{aq50}-\eqref{aq54}, we obtain
\begin{equation}
\frac{\partial z}{\partial \Psi}=\chi\left(\frac{\|\bar{\mathbf{q}}\|\sin(\pi-\angle AEB)\sin\angle ABE}{\|\mathbf{q}_e\|\sin\angle BAE\sin(\pi-\angle AEB)}\alpha-1\right).
\end{equation}
From the sine theorem, we know
\begin{equation}
\frac{\|\bar{\mathbf{q}}\|}{\sin\angle BAE}=\frac{\|\mathbf{q}_e\|}{\sin\angle ABE}
\end{equation}
and thus
\begin{equation}\label{aq57}
\frac{\partial z}{\partial \Psi}=\chi(\alpha-1)\geq0.
\end{equation}
Substituting \eqref{q20} into \eqref{aq49}, we have
\begin{align}\label{aq58}
\chi=&\frac{\|\mathbf{q}_b\|^2}{\|\mathbf{q}_e\|}\cdot\frac{\left(\mathbf{q}_a(\theta)^T\mathbf{q}_e\right)^2}{\left(\mathbf{q}_a(\theta)^T\mathbf{q}_b\right)^2\|\mathbf{q}_a(\theta)\|}\nonumber\\
=&\frac{\|\mathbf{q}_e\|}{\|\mathbf{q}_a(\theta)\|}\cdot\frac{\cos^2\angle QAE}{\cos^2\angle QAB}.
\end{align}
Combining \eqref{aq57} and \eqref{aq58}, we obtain Proposition 6.

\section{Proof of Proposition 7}

Under this condition of \eqref{cq43}, $\Upsilon_1$ is a monotonically increasing affine function with respect to $\alpha$.
 
Taking the first-order partial derivative of $\Upsilon_2$ with respect to $\alpha$ yields
\begin{align}
\frac{\partial \Upsilon_2}{\partial \alpha}=\frac{M(\|\bar{\mathbf{q}}\|^2\alpha+\bar{\mathbf{q}}^T\mathbf{q}_e)}{\sqrt{\|\bar{\mathbf{q}}\|^2\alpha^2+2\bar{\mathbf{q}}^T\mathbf{q}_e \alpha+\|\mathbf{q}_e\|^2}}
\end{align}
where $M=\frac{\gamma L_b L_e G_0}{\|\mathbf{q}_b\|\|\mathbf{q}_e\|}$. Taking the second-order partial derivative of $\Upsilon_2$ with respect to $\alpha$ yields
\begin{align}
\frac{\partial^2 \Upsilon_2}{\partial \alpha^2}=\frac{M(\|\bar{\mathbf{q}}\|^2\|\mathbf{q}_e\|^2-(\bar{\mathbf{q}}^T\mathbf{q}_e)^2)}{(\|\bar{\mathbf{q}}\|^2\alpha^2+2\bar{\mathbf{q}}^T\mathbf{q}_e \alpha+\|\mathbf{q}_e\|^2)^{3/2}}.
\end{align}
Because $\|\bar{\mathbf{q}}\|^2\|\mathbf{q}_e\|^2-(\bar{\mathbf{q}}^T\mathbf{q}_e)^2\geq0$, $\Upsilon_2$ is a convex function with respect to $\alpha$. The value of $\alpha$ which maximizes $\Upsilon_2$ is $\alpha^\star$ defined in \eqref{cq63}.

We consider three cases, i.e., $\zeta_2>0$, $\zeta_2<0$, and $\zeta_2=0$. 

When $\zeta_2>0$ and thus
\begin{equation}
L_b\|\mathbf{q}_e\|-L_e\|\mathbf{q}_b\|>\gamma L_b L_e G_0\|\bar{\mathbf{q}}\|,
\end{equation}
we have
\begin{align}
\frac{\partial \Upsilon_1}{\partial \alpha}>\lim_{\alpha\rightarrow\infty}\frac{\partial \Upsilon_2}{\partial \alpha}=M\|\bar{\mathbf{q}}\|=\frac{\gamma L_b L_e G_0\|\bar{\mathbf{q}}\|}{\|\mathbf{q}_b\|\|\mathbf{q}_e\|}.
\end{align}
Therefore, $\Upsilon_1$ and $\Upsilon_2$ have one intersection point, which corresponds to one solution of equation \eqref{cq45}. The other solution is obtained from $-\Upsilon_1=\Upsilon_2$. Because of $\Upsilon_2>0$ and the monotonically increasing $\Upsilon_1$, from Fig. 6(a), we know that the optimal $\alpha$ is \eqref{cq62}.

\begin{figure}
\centering
\includegraphics[width=3.4in]{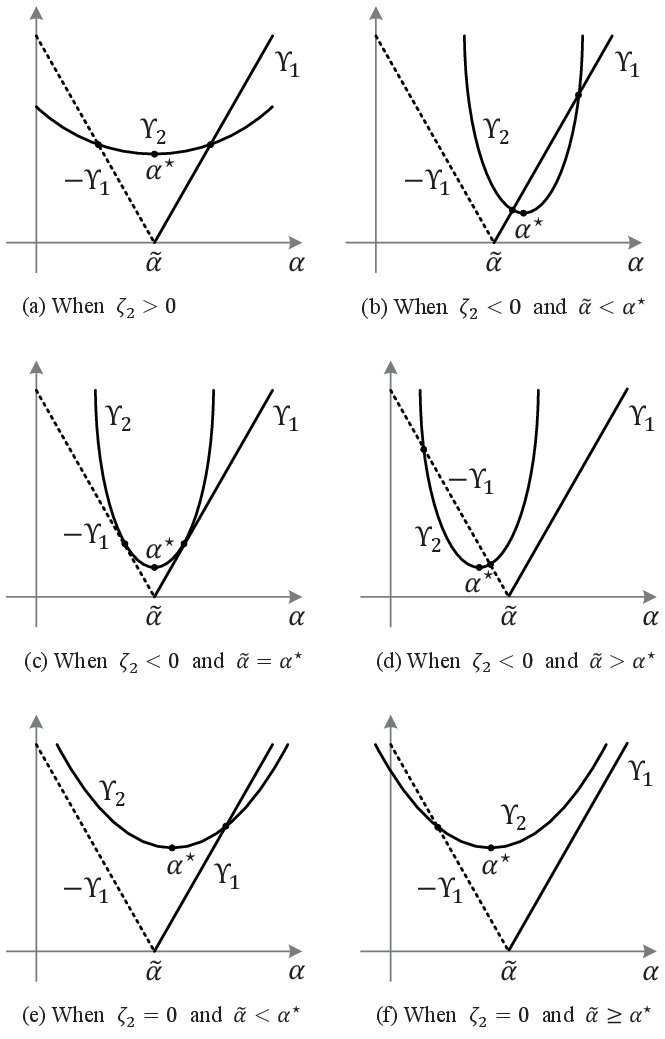}
\caption{The relationship between $\Upsilon_1$ and $\Upsilon_2$.}
\end{figure}

When $\zeta_2<0$, we have
\begin{align}
\frac{\partial \Upsilon_1}{\partial \alpha}<\lim_{\alpha\rightarrow\infty}\frac{\partial \Upsilon_2}{\partial \alpha}.
\end{align}
Denote the solution to $\Upsilon_1=0$ as $\tilde{\alpha}$ shown in \eqref{cq63}.

If $\tilde{\alpha}<\alpha^\star$, Fig. 6(b) shows that $\Upsilon_1=\Upsilon_2$ has two distinct solutions, i.e., 
\begin{align}
\alpha_1&=\frac{-\zeta_1+\sqrt{\zeta_1^2-4\zeta_2\zeta_0}}{2\zeta_2},\\
\alpha_2&=\frac{-\zeta_1-\sqrt{\zeta_1^2-4\zeta_2\zeta_0}}{2\zeta_2}.
\end{align}
Since $\alpha_1<\alpha_2$, we know that 
\begin{align}
\frac{\partial C^{\text{LoS}}_s(\alpha)}{\partial \alpha}>0 \text{ for } \alpha<\alpha_1 \text{ and }
\frac{\partial C^{\text{LoS}}_s(\alpha)}{\partial \alpha}<0 \text{ for } \alpha>\alpha_1.
\end{align}
This shows that $\alpha_1$ achieves the maximum. Similarly, we can prove that $\alpha_2$ achieves the minimum. Therefore, we know that the optimal $\alpha$ is \eqref{cq62}.

If $\tilde{\alpha}=\alpha^\star$, Fig. 6(c) shows that $\Upsilon_1=\Upsilon_2$ has a solution $\alpha_1$ and $-\Upsilon_1=\Upsilon_2$ has another solution $\alpha_2$. Under this condition, we know $\Upsilon_2\geq\Upsilon_1$. From \eqref{cq42}, we have $\frac{\partial C^{\text{LoS}}_s(\alpha)}{\partial \alpha}\geq0$. The optimal $\alpha$ is $\alpha^o=\alpha_{\max}$.

If $\tilde{\alpha}>\alpha^\star$, Fig. 6(d) shows that $-\Upsilon_1=\Upsilon_2$ has two distinct solutions whereas $\Upsilon_1=\Upsilon_2$ has no solution. Furthermore, because of $\frac{\partial C^{\text{LoS}}_s(\alpha)}{\partial \alpha}>0$, the optimal $\alpha$ is $\alpha^o=\alpha_{\max}$.

When $\zeta_2=0$, we have
\begin{align}
\frac{\partial \Upsilon_1}{\partial \alpha}=\lim_{\alpha\rightarrow\infty}\frac{\partial \Upsilon_2}{\partial \alpha}.
\end{align}
If $\tilde{\alpha}<\alpha^\star$, Fig. 6(e) shows that $\Upsilon_1=\Upsilon_2$ has a solution, which achieves the maximum value of $C^{\text{LoS}}_s(\alpha)$. Therefore, we know that the optimal $\alpha$ is \eqref{cq65}.

If $\tilde{\alpha}\geq\alpha^\star$, Fig. 6(f) shows that $\frac{\partial C^{\text{LoS}}_s(\alpha)}{\partial \alpha}>0$ over $1\leq\alpha\leq\alpha_{\max}$ and thus the optimal $\alpha$ is $\alpha^o=\alpha_{\max}$.


\begin{thebibliography}{10}

\bibitem{KKWong21} K.-K. Wong, A. Shojaeifard, K.-F. Tong, and Y. Zhang, ``Fluid antenna systems," \emph{IEEE Trans. Wireless Commun.}, vol. 20, no. 3, pp. 1950--1962, Mar. 2021.
    
\bibitem{TWu24} T. Wu et al., ``Fluid antenna systems enabling 6G: Principles, applications, and research directions," 2024, \emph{arXiv:2412.03839.} [Online]. Available: http://arxiv.org/abs/2412.03839
    
\bibitem{TWu25} T. Wu et al., ``Scalable FAS: A new paradigm for array signal processing," 2025, \emph{arXiv:2508.10831.} [Online]. Available: http://arxiv.org/abs/2508.10831

\bibitem{JZheng24} J. Zheng, T. Wu, X. Lai, C. Pan, M. Elkashlan, and K.-K. Wong, ``FAS-assisted NOMA short-packet communication systems," IEEE Trans. Veh. Technol., vol. 73, no. 7, pp. 10732-10737, Jul. 2024.     

\bibitem{LZhu24} L. Zhu, W. Ma, and R. Zhang, ``Modeling and performance analysis for movable antenna enabled wireless communications," \emph{IEEE Trans. Wireless Commun.}, vol. 23, no. 6, pp. 6234--6250, Jun. 2024. 
    
\bibitem{XShao25} X. Shao, Q. Jiang, and R. Zhang, ``6D movable antenna based on user distribution: Modeling and optimization," \emph{IEEE Trans. Wireless Commun.}, vol. 24, no. 1, pp. 355--370, Jan. 2025.

\bibitem{XShao252} X. Shao, R. Zhang, Q. Jiang, and R. Schober, ``6D movable antenna enhanced wireless network via discrete position and rotation optimization," \emph{IEEE J. Sel. Areas Commun.}, vol. 43, no. 3, pp. 674--687, Mar. 2025.

\bibitem{Ghadi24} F. R. Ghadi, K. -K. Wong, F. J. L\'opez-Mart\'inez, W. K. New, H. Xu, and C.-B. Chae, ``Physical layer security over fluid antenna systems: Secrecy performance analysis," \emph{IEEE Trans. Wireless Commun.}, vol. 23, no. 12, pp. 18201--18213, Dec. 2024.

\bibitem{Sanchez24} J. D. Vega-S\'anchez, L. F. Urquiza-Aguiar, H. R. C. Mora, N. V. O. Garz\'on, and D. P. M. Osorio, ``Fluid antenna system: Secrecy outage probability analysis," \emph{IEEE Trans. Veh. Technol.}, vol. 73, no. 8, pp. 11458--11469, Aug. 2024.

\bibitem{JYao25} J. Yao, L. Xin, T. Wu, M. Jin, K.-K. Wong, C. Yuen, and H. Shin, ``FAS for secure and covert communications," \emph{IEEE Internet Things J.}, vol. 12, no. 11, pp. 18414--18418, Jun. 2025.

\bibitem{GHu24} G. Hu, Q. Wu, K. Xu, J. Si, and N. Al-Dhahir, ``Secure wireless communication via movable-antenna array," \emph{IEEE Signal Process. Lett.}, vol. 31, pp. 516--520, Jan. 2024.

\bibitem{JTang25} J. Tang, C. Pan, Y. Zhang, H. Ren, and K. Wang, ``Secure MIMO communication relying on movable antennas," \emph{IEEE Trans. Commun.}, vol. 73, no. 4, pp. 2159--2175, Apr. 2025.

\bibitem{JDing25} J. Ding, Z. Zhou, and B. Jiao, ``Movable antenna-aided secure full-duplex multi-user communications," \emph{IEEE Trans. Wireless Commun.}, vol. 24, no. 3, pp. 2389--2403, Mar. 2025.

\bibitem{YQian} Y. Qian, J. Jiang, J. Ding, X. Dan, and H. Chu, ``6DMA-assisted secure wireless communications," 2025, \emph{arXiv:2509.16698.} [Online]. Available: http://arxiv.org/abs/2509.16698

\bibitem{BZheng25} B. Zheng, Q. Wu, and R. Zhang, ``Rotatable antenna enabled wireless communication: Modeling and optimization," 2025, \emph{arXiv:2501.02595.} [Online]. Available: http://arxiv.org/abs/2501.02595
    
\bibitem{BZheng252} B. Zheng, T. Ma, C. You, J. Tang, R. Schober, and R. Zhang, ``Rotatable antenna enabled wireless communication and sensing: Opportunities and challenges," 2025, \emph{arXiv:2505.16828.} [Online]. Available: http://arxiv.org/abs/2505.16828
    
\bibitem{QWu25} Q. Wu, B. Zheng, T. Ma, and R. Zhang, ``Modeling and optimization for rotatable antenna enabled wireless communication," in \emph{Proc. Int. Conf. Commun. (ICC), 2025}, pp. 1055-1060.

\bibitem{RZhao25} R. Zhao, Y. Xu, S. Yang, H. Chen, and C. Assi, ``Beamforming for movable and rotatable antenna enabled multi-user communications," 2025, \emph{arXiv:2503.11130.} [Online]. Available: http://arxiv.org/abs/2503.11130

\bibitem{XXiong25} X. Xiong, B. Zheng, W. Wu, X. Shao, L. Dai, M.-M. Zhao, and J. Tang, ``Efficient channel estimation for rotatable antenna-enabled wireless communication," \emph{IEEE Wireless Commun. Lett.}, 2025,  doi:10.1109/LWC.2025.3601979.

\bibitem{XPeng25} X. Peng, Q. Wu, Z. Zheng, W. Chen, Y. Zhu, and Y. Gao, ``Rotatable antenna enabled spectrum sharing: Joint antenna orientation and beamforming design," 2025, \emph{arXiv:2509.19912.} [Online]. Available: http://arxiv.org/abs/2509.19912

\bibitem{XZhang25} X. Zhang, L. Xiang, J. Wang, X. Gao, D. W. K. Ng, and R. Schober, ``Rotatable antenna array enabled UAV mmWave massive MIMO communication," \emph{IEEE Trans. Commun.}, 2025, doi:10.1109/TCOMM.2025.3622962.

\bibitem{CZhou25} C. Zhou, C. You, B. Zheng, X. Shao, and R. Zhang, ``Rotatable antennas for integrated sensing and communications," \emph{IEEE Wireless Commun. Lett.}, vol. 14, no. 9, pp. 2838--2842, Sep. 2025.

\bibitem{LDai25} L. Dai, B. Zheng, Q. Wu, C. You, R. Schober, and R. Zhang, ``Rotatable antenna-enabled secure wireless communication," \emph{IEEE Wireless Commun. Lett.}, 2025, doi:10.1109/LWC.2025.3593258. 


\bibitem{Bhargav} N. Bhargav, S. L. Cotton, and D. E. Simmons, ``Secrecy capacity analysis over $\kappa$-$\mu$ fading channels: Theory and applications," \emph{IEEE Trans. Commun.}, vol. 64, no. 7, pp. 3011-3024, Jul. 2016.

\bibitem{Abramowitz} M. Abramowitz and I. A. Stegun, \emph{Handbook of Mathematical Functions With Formulas, Graphs and Mathematical Tables}. New York, NY, USA: Dover, 1964.

\bibitem{Avriel} M. Avriel, W. E. Diewert, S. Schaible, and I. Zang, \emph{Generalized Concavity}, New York, NY, USA: Plenum Press, 1988.
    
\bibitem{Segura} J. Segura, ``Monotonicity properties for ratios and products of modified Bessel functions and sharp trigonometric bounds," \emph{Results Math.} vol. 76, no. 221, pp. 1-22, 2021.

\end{thebibliography}
\end{document}